\documentclass[final,3p,times]{elsarticle}

\pdfoutput=1
\usepackage{amssymb}
\usepackage{url}
\usepackage{natbib}
\usepackage{amsthm}
\usepackage{multirow}
\usepackage[skip=6pt plus1pt, indent=0pt]{parskip}
\usepackage[none]{hyphenat}
\usepackage{caption} 
\usepackage[labelfont=bf]{caption}
\biboptions{sort&compress}
\journal{Applied Energy}

\begin{document}
\emergencystretch 3em

\begin{frontmatter}



\title{Realising large areal capacities in liquid metal batteries: a battery design concept for mass transfer enhancement}


\author[label1,label2]{Declan Finn Keogh}
\author[label2]{Mark Baldry}
\author[label1]{Victoria Timchenko}
\author[label1]{John Reizes}
\author[label1]{Chris Menictas}

\affiliation[label1]{organization={School of Mechanical and Manufacturing Engineering, UNSW},
            city={Sydney},
            postcode={2052}, 
            state={NSW},
            country={Australia}}

\affiliation[label2]{organization={School of Biomedical Engineering, The University of Sydney},
            city={Sydney},
            postcode={2008}, 
            state={NSW},
            country={Australia}}

\begin{abstract}
Liquid metal batteries (LMBs) are a promising grid-scale storage device however, the scalability of this technology and its electrochemical performance is limited by mass transport overpotentials. In this work, a numerical model of a three-layer LMB was developed using a multi-region approach. An alternative design concept for the battery aimed at reducing mass transport overpotentials, increasing cell capacity, and improving electrochemical cell performance was implemented and evaluated. The design consisted of a coil implanted in the cathode, which induced mixing in the layer. Four cases were compared: three in a 241 Ah LMB at 0.3, 0.5 and 1 A/cm$^{2}$, and one in a larger 481 Ah LMB at 0.5 A/cm$^{2}$. LMB performance was determined by comparison against baseline diffusion cases and a change in molar fraction of 0.1. The modified LMB exhibited dramatic performance increases with a 78\% and 85\% reduction in mass-transport overpotentials at 0.3 A/cm$^{2}$ and 0.5 A/cm$^{2}$, respectively. The improved performance of the battery was directly attributed to the flow generated in the cathode. It was found that the coil substantially increased the poloidal volumetric average velocity. Periodically, vortices formed that removed concentration gradients from the cathode-electrolyte interface, minimising concentration polarisation. The viability of the design was tested in a lab-scale prototype using Galinstan as the working fluid. The velocity of the induced flow was determined using particle image velocimetry (PIV), and the results compared to the numerical model. There was a close match between the experimental and numerical results, validating the numerical model and the viability of the design. Implementation of this design concept in future LMBs could lead to the realisation of extended discharge capacities and improved voltages. Future work is planned to test the coil in a working battery.
\end{abstract}

\begin{graphicalabstract}
\begin{figure}[htp]
    \centering
    \includegraphics[height=5cm]{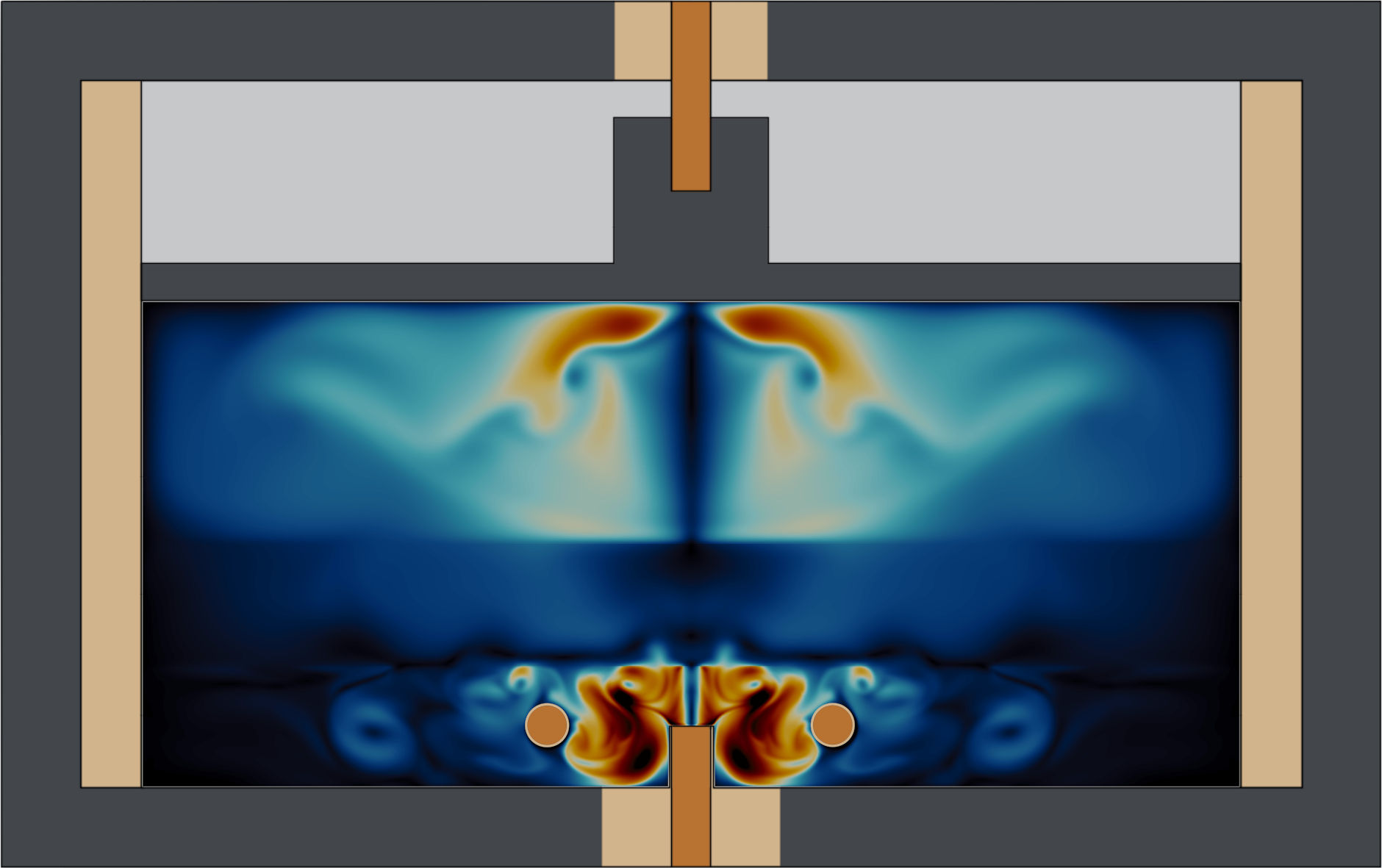}
    \label{fig:diagram1}
\end{figure}
\end{graphicalabstract}

\begin{highlights}
\item Design concept for mass transfer enhancement
\item A small coil is used to generate a mixing electro-vortex flow
\item The coil minimally increases ohmic overpotentials
\item Large areal capacities with minimal mass transport overpotentials are achieved
\item PIV results of the liquid metal flow in a prototype validate the numerical model
\end{highlights}

\begin{keyword}
Liquid metal battery \sep Mass transport overpotential \sep Concentration overpotential \sep Mixing \sep Mass transfer \sep Mass transport \sep Electro-vortex flow \sep Swirling electro-vortex flow \sep Mass transfer enhancement


\end{keyword}

\end{frontmatter}


\section{Introduction}
\label{Introduction}
Addressing the challenges of global warming requires the decarbonisation of global electricity grids. This energy transition is occurring worldwide via the uptake of renewables, especially wind and solar; in 2021 alone, wind and solar generation capacity grew by 226 GW and reached a 10.2$\%$ share of electricity generation worldwide \cite{BritishPetroleum2022Statistical2022}. While the rapid transition to renewables is necessary, these energy sources present challenges to modern electricity grids. Renewable energy generation is, by nature, intermittent, and peak power generation does not necessarily align with peak power demand \cite{Leadbetter2012SelectionElectricity}. This intermittency necessitates energy storage with durations anywhere from minutes to hours and days to secure the electricity grid. Batteries are well-positioned to supply this storage with a range of discharge durations depending on the technology.
\par
Liquid metal batteries (LMBs) are a grid-scale energy storage technology developed to enable this transition from carbon-intensive energy sources to renewables. The all-liquid metal battery is typically comprised of a less dense electropositive anode, a molten salt electrolyte, and a dense electronegative cathode. The large differences in density between the immiscible layers result in a gravity-stratified system that does not require any physical separators. This creates cost efficiencies as the cell design is simple and easy to manufacture; typically, the electroactive materials are housed in a grade 304L stainless-steel cylinder sealed with feed-throughs for the current collectors \cite{Kim2013LiquidFuture}. Beyond low cell construction costs, the batteries also have long cycle lifetimes, low material costs, and high-power densities. The Li${||}$Sb-Pb chemistry has been shown to have fade rates as low as 0.002${\%}$ per cycle \cite{Ouchi2017PositiveBattery} and low electrode material costs of {\$}65 USD/kWh \cite{Wang2014Lithium-antimony-leadStorage}. The Li${||}$Te chemistry has power densities of 495 Wh/kg with high discharge voltages of 1.6 V \cite{Li2018Tellurium-tinApplications}. Additionally, most chemistries have been shown to be able to discharge at current densities up to 1 A/cm$^{2}$ due to the low resistances of the metal electrodes and low activation overpotentials \cite{Wang2014Lithium-antimony-leadStorage,Li2016HighElectrode,Ning2015Self-healingStorage,Dai2018CapacityBattery,Xie2020High-performanceBattery,Newhouse2017Charge-TransferElectrodes}. Owing to these impressive performance characteristics, the technology has gained interest in the field due to its immense promise as a grid-scale storage device.
\par
\begin{figure}[htp]
    \centering
    \includegraphics[width=15cm]{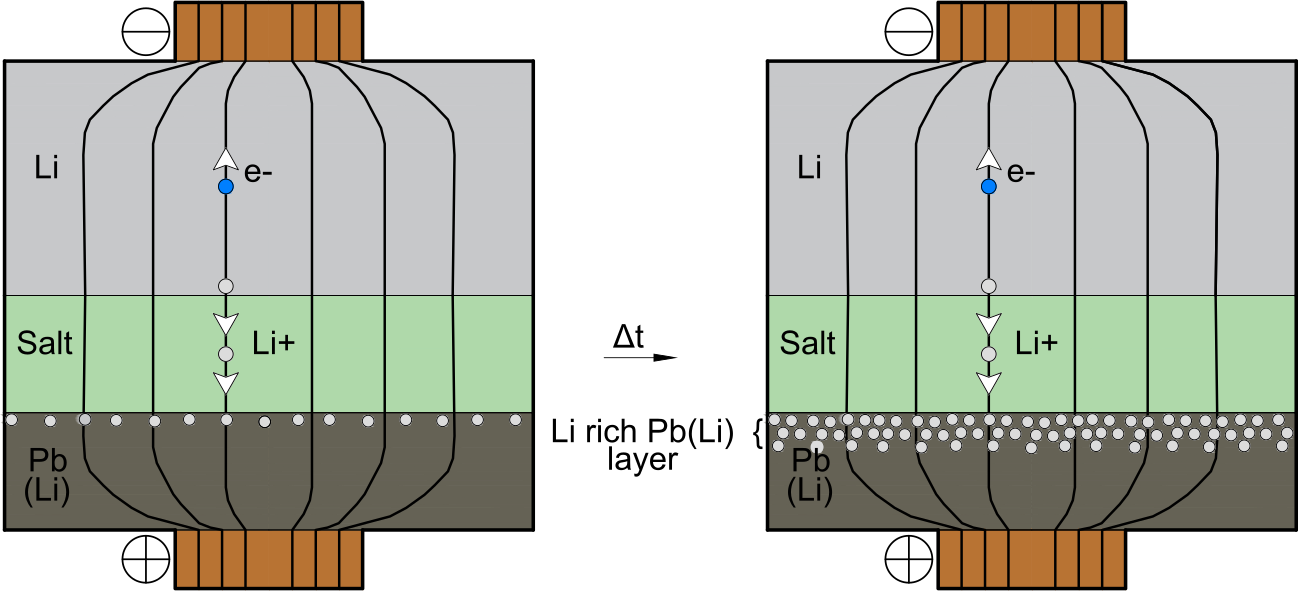}
    \caption{Discharge process in a Liquid Metal Battery. Anodic ions are transported through the electrolyte layer into the cathode. Once there, the anodic atoms accumulate near the cathode-electrolyte interface.}
    \label{fig:diagram2}
\end{figure}
\par
However, LMBs still have developmental challenges that must be overcome if they are to achieve the stability, robustness, and discharge capacity to be utilised for energy services such as energy arbitrage and load-shifting. The main obstacle for this technology is the mass transport overpotential that occurs during discharge. Mass transport overpotential can have significant impacts on a cell's performance, and these can be readily observed in the literature, as summarised in Table 1.
\begin{table}[ht]
\caption{Experimental observations of mass transport overpotentials.}
\label{tab:literature}
\centering
\begin{tabular}{ p{2cm} p{11cm} p{2cm} }
 \hline
 \hline
 Chemistry& Observation(s)& Reference \\
 \hline
 Ca${||}$Bi   &At 25 mA/cm$^{2}$ the battery discharged to a molar fraction, x, of ${x_\mathrm{Ca}=0.24}$, at 150 mA/cm$^{2}$ this was reduced ${x_\mathrm{Ca}= 0.21}$, at 200 mA/cm$^{2}$ this was reduced to an ${x_\mathrm{Ca}=0.17}$. An analysis of the losses in the battery showed that 59${\%}$ were from mass transport overpotentials.    &\cite{Kim2013CalciumbismuthBatteries}\\
 Li${||}$Sb,Sn &A 192 Ah cell had a maximum anode capacity utilisation of 94${\%}$ while a 400 Ah cell had a maximum anode capacity utilisation of 85${\%}$ even at low discharge rates. &\cite{Zhou2022IncreasingBattery}\\
 Li${||}$Bi &As the discharge current density increased from 200 mA/cm$^{2}$ to 1250 mA/cm$^{2}$ a 30${\%}$ reduction in the reported capacity was observed. &\cite{Ning2015Self-healingStorage}\\
 Li${||}$Bi,Sb    &A small capacity battery (147 mAh) could only be discharged to 50${\%}$ of the reported capacity at 1000 mA/cm$^{2}$. &\cite{Dai2018CapacityBattery}\\
 \hline
 \hline
\end{tabular}
\end{table}
\vspace{-1pc}
\par
Mass transport overpotentials can cause the battery to discharge at lower voltages, limiting the performance of LMBs. Reduced discharge voltages also cause a coupled reduction in capacity utilisation as the battery voltage falls below the cut-off voltage to stop discharging before the battery reaches its theoretical capacity. Multiple literature reviews on the suitability of batteries for grid-scale storage identify discharge time as the most important determinant for a battery technology's commercial success \cite{Leadbetter2012SelectionElectricity,Rahman2020AssessmentReview,Comello2019TheStorage}. The reduction in capacity utilisation of LMBs in Table 1 as the capacity, and discharge time, are increased is seemingly a barrier to this technology’s commercial competitiveness in the grid-scale market. This paper will focus on addressing mass transport overpotentials in grid-scale LMBs with large capacities.
\par
Mass transport overpotentials in the battery are caused by concentration polarisation between the cathode-electrolyte interface and the bulk of the cathode. During discharge, ions migrate from the anode through the electrolyte layer and enter the cathode through the cathode-electrolyte interface forming an alloy in the cathode. The anodic atoms are then transported from the interface by diffusion into the bulk of the cathode \cite{Personnettaz2019MassBatteries,Ashour2018Convection-diffusionBatteries}. Due to the high discharge current achievable by LMBs, diffusion of species away from the interface occurs at a slower rate than the rate of atoms entering the cathode. This causes the concentration of anodic atoms at the cathode-electrolyte interface to be greater than the volumetric concentration in the cathode as shown in Fig. 1. This concentration polarisation causes the mass transport overpotentials in the battery.
\par
The key to solving this issue is to homogenise the concentration of anodic atoms in the cathode. Owing to the all-liquid composition of the batteries, flows can be generated in the cathode to mix the atoms and extend the operating capacity of the batteries \cite{Ashour2018Convection-diffusionBatteries,Weber2020NumericalFlow,Herreman2020SolutalBatteries,Herreman2021EfficientBatteries}. However, a problematic aspect of these concentration polarisations is that they cause significant density stratifications in the cathode \cite{Herreman2020SolutalBatteries}. The anode metal must be less dense than both the electrolyte layer and the cathode for the liquids to gravity stratify. This means that the alloy at the cathode-electrolyte interface has a much lower density than the rest of the cathode. The induced flow must be vigorous enough to break through this density stratification and mix the cathode \cite{Herreman2020SolutalBatteries}.
\par
There are several sources of fluid motion in the battery, including Rayleigh-Bénard convection \cite{Kollner2017ThermalModel,Personnettaz2018ThermallyBatteries, Personnettaz2022LayerBatteries,Keogh2023ModellingBatteries, Cheng2022Oscillations1.4-3}, electro-vortex flow \cite{Keogh2023ModellingBatteries,Herreman2019NumericalBatteries,Herreman2020SolutalBatteries,Keogh2021ModellingBatteries,Ashour2018CompetingBatteries,Weber2018ElectromagneticallyBatteries} and swirl flow \cite{Keogh2023ModellingBatteries,Herreman2021EfficientBatteries,Weber2020NumericalFlow}. While Rayleigh-Bénard convection will be present throughout discharge, the flow that is viscously coupled throughout the battery does not impact concentration stratification in the cathode \cite{Personnettaz2018ThermallyBatteries,Personnettaz2022LayerBatteries,Keogh2023ModellingBatteries}. For this reason, the rest of this discussion will be limited to electro-vortex flow and swirl flow. Electro-vortex flow is induced by the Lorentz force created from flux concentrations of the current density at the current collectors. Specifically, the current in the battery interacts with its own magnetic field to generate a Lorentz force \cite{Shercliff1970FluidSource,Bojarevics1989ElectricallyFlows,Weber2015TheBatteries,Herreman2019NumericalBatteries}. The resultant flow consists of an axially oriented jet that expels fluid away from the current collector, penetrating the buoyant and concentration stratified layer reducing concentration polarisation \cite{Herreman2020SolutalBatteries}. However, during discharge, the growing density stratification suppresses electro-vortex flow, and, depending on the dimensions of the battery, it can be expected not to act for the entire length of discharge \cite{Herreman2020SolutalBatteries,Keogh2023ModellingBatteries}. Clearly, this is not ideal for a LMB with an extended discharge capacity.
\par
A more promising flow to mix the cathode and extend the discharge capacity of LMBs is swirl flow. Swirl flow is caused by the Lorentz force generated by the interaction of an externally applied vertical magnetic field with the internal current. The external vertical magnetic field and horizontal current flux in the liquid metal layers create a rotational Lorentz force. This rotational force changes the previously poloidal electro-vortex flow to a toroidally dominant flow \cite{Davidson1999TheForces}. The induced flows can be vigorous and are controlled by increasing and decreasing the magnitude of the external magnetic field, \textbf{\textit{B}}, or conversely by increasing and decreasing the discharge current density, \textbf{\textit{J}} \cite{Herreman2021EfficientBatteries}. It has been shown this flow is more effective at reducing mass transport overpotentials in the battery \cite{Weber2020NumericalFlow}, and that it is resistant to stratifying buoyancy forces in the cathode \cite{Keogh2023ModellingBatteries}. Theoretically, a strong enough flow can be induced to completely homogenise concentration in the cathode \cite{Herreman2021EfficientBatteries}. Our aim here, is to introduce a grid-scale capacity LMB with a viable design to realise a swirl flow of an intensity sufficient to mix the cathode layer.
\section{Design Concept}
\label{Design Concept}
There were several design constraints set before designing the modified LMB. These include:
\par
 \begin{enumerate}
 \itemsep6pt
     \item In a typical battery module, individual cells are placed such that their exterior conducting walls are in contact \cite{BradwellUS20210043982A1Patents}. This has the advantage of allowing the batteries to operate in parallel. It also allows the battery module to be maintained at a stable temperature. Guo \textit{et al.} has shown that air gaps cannot be present in the battery as the low thermal conductivity of air makes the temperature in the battery unstable \cite{Guo2018NumericalModule}. The batteries must be able to be placed in such a manner that the vessel walls are still touching.
     \item In a battery module, LMBs are vertically stacked \cite{BradwellUS20210043982A1Patents}. This is achieved either using conductive steel shelves that the batteries can electrically contact or by using a coupling of the current collectors. The design must still allow for vertical stacking.
     \item The proposed design must limit increases to ohmic overpotentials, a major source of losses in LMBs \cite{Kim2013CalciumbismuthBatteries}.
     \item The design must work for all chemistries regardless of the density differences.
 \end{enumerate}
 \vspace{-0.5pc}
To meet these design constraints, the proposed design consists of a three-layer LMB housed inside a cylindrical steel vessel made from grade 304L stainless steel. The wall thickness is 6.5 mm and the inner radius is 0.5\textit{D} = \textit{R} = 50.14 mm. The cathode is electrically isolated from the anode by a hollow Al$_{2}$O$_{3}$ cylinder with a wall thickness of 5 mm. The current collectors are also made from stainless steel with a copper wire in electrical contact with the electrodes via a feed-through in the vessel. The electrolyte layer is 10 mm high, a standard height used in the literature \cite{Ning2015Self-healingStorage,Zhou2022IncreasingBattery,Weber2020NumericalFlow,Herreman2020SolutalBatteries,Herreman2021EfficientBatteries,Kollner2017ThermalModel,Zhou2022Multi-fieldPerformance}. These aspects of the design are similar to the literature \cite{Ning2015Self-healingStorage}.
\par
\begin{figure}[htp]
    \centering
    \includegraphics[width=15cm]{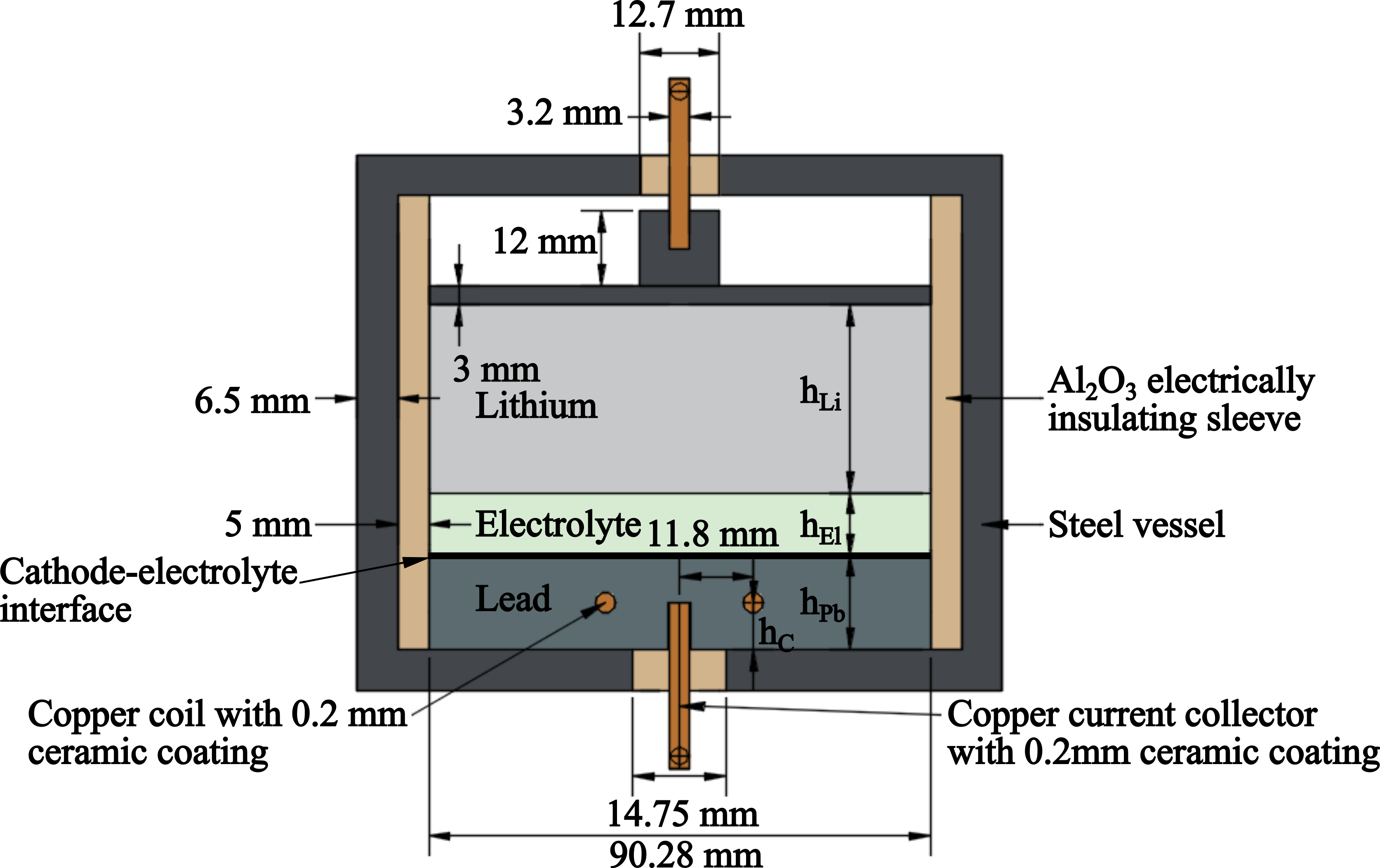}
    \caption{Cell concept used for the numerical studies in this work. It features a coil embedded inside the cathode layer to generate a vertical magnetic field and a current collector raised inside the cathode layer to half the height of the cathode.}
    \label{fig:diagram3}
\end{figure}
\par
The changes consist of a coil embedded inside the cathode to generate a vertical magnetic field to induce a swirl flow, see Fig. 2. Embedding the coil inside the cathode layer eliminates interference with any existing components of a battery module. The coil is connected in series to the battery allowing the battery’s own discharge current to create the magnetic field.  A strong field is generated using a single small loop of copper wire which minimises the ohmic overpotentials from the additional wiring. The coil is embedded at half the height of the cathode layer; therefore, the maximum vertical magnetic field is also generated at this height. The current collector inside the battery is raised to this height so that the current exiting the current collector is exposed to this field. Finally, the coil and current collector are insulated from the cathode where required using a ceramic coating.
\section{Methods}
\label{Methods}
There are several challenges to evaluating a design change such as the one proposed here in a working LMB. The opaque working fluid (metal) prevents the use of flow visualisation techniques such as PIV in the battery, and the high operating temperature prevents the use of ultrasound flow mapping. There are experimental results at temperatures up to 160\textdegree C \cite{Perez2015UltrasoundElectrode,Ashour2018CompetingBatteries}, but beyond this temperature, a waveguide is required \cite{Eckert2003VelocityGuide}. LMBs typically operate at temperatures between 450-500\textdegree C \cite{Kim2013LiquidFuture}, making numerical modelling an ideal method to evaluate the proposed design changes.
\par
Numerical modelling is still challenging, as the battery consists of three layers, and three-dimensional (3D) three-phase models are extremely computationally expensive. It would be challenging to compute processes that represent even a fraction of the discharge time of a battery using this approach. Therefore, in this work, a two-dimensional (2D) axisymmetric model, using a technique introduced by Weber \textit{et al.} \cite{Weber2020NumericalFlow} and Personnettaz \textit{et al.} \cite{Personnettaz2022LayerBatteries}, was developed to evaluate the proposed design. A lab-scale prototype was constructed and tested using a layer of Galinstan and a seeded layer of HCl, and the resulting flow was analysed using PIV.
\subsection{Numerical study design}
\label{Numerical study design}
The chemistry chosen to test the design was the lithium-lead chemistry, Li${||}$Pb. This chemistry has the strongest density stratification owing to the low density of lithium, 487 kg/m$^{3}$, and the high density of lead, 10,430 kg/m$^{3}$. Accordingly, if the flow is able to mix the cathode for this anode and cathode pairing, then it will work for all chemistries. The material properties of the pure layers are also well defined at the operating temperature of the cell, ${T=500}$\textdegree C, as well as the material properties of the cathode alloy at different molar fractions of lithium. This is because the alloy lithium-lead, indicated as Pb(Li), is a candidate for tritium breeder blankets for nuclear fusion reactors \cite{Martelli2019LiteratureProperties,MasdelesValls2008LeadlithiumTechnology,Khairulin2017InterdiffusionMelts,Schulz1991ThermophysicalLi17Pb83alloy,Khairulin2017VolumetricMelts}. In all cases, the material properties used are for Pb(Li) at a molar fraction of ${x_{\mathrm{Li}}=0.25}$. This intermediate molar fraction was used as between ${0.17<x_{\mathrm{Li}}<0.6}$ the density of the cathode alloy varies linearly with increasing concentrations of lithium \cite{Herreman2020SolutalBatteries,Khairulin2017VolumetricMelts}. This means that the Boussinesq approximation can be used to capture the changes in density during discharge. The electrolyte modelled was LiF-LiCl-LiI (3.2-13-83.8) which has been used in similar cells and is compatible with the chemistry \cite{Kim2013LiquidFuture}. The electrical conductivities, ${\sigma}$, of steel and copper are required for computation of the electric and magnetic fields, and at ${T=500}$\textdegree C they are 19,790,000 and 1,370,000 S/m, respectively \cite{Matula1979ElectricalSilver,Kim1975ThermophysicalK}. The material properties of the layers are provided in Table 2.


\begin{table}[ht]
\caption{Material properties at \textit{T} = 500\textdegree C}
\centering
\begin{tabular}{llll}
\hline
\hline
\textbf{Component}                                                                                         & \textbf{Material Property}                                                  & \textbf{Value} & \textbf{Reference}                 \\ \hline
\multirow{6}{*}{Lithium Anode}                                                                             & Density ${\rho}$ (kg/m$^{3}$)                                         & 489            & \multirow{6}{*}{{}\cite{Zinkle1998SummaryCoolants}{}}          \\
                                                                                                           & Volumetric thermal expansion coefficient ${\beta_{\mathrm{Th}}}$ (1/K) & 2.05x10$^{-4}$      &                                    \\
                                                                                                           & Kinematic viscosity \textit{v} (m$^{2}$/s)                                                & 6.72x10$^{-7}$     &                                    \\
                                                                                                           & Heat capacity ${c_{\mathrm{p}}}$ (J/kgK)                                                  & 4,210          &                                    \\
                                                                                                           & Thermal conductivity ${\lambda}$ (W/mK)                          & 53.5           &                                    \\
                                                                                                           & Electrical conductivity ${\sigma}$ (S/m)           & 2,620,000       &                                    \\ \hline
\multirow{6}{*}{\begin{tabular}[c]{@{}l@{}}LiF-LiCl-LiI \\    \\ (3.2-13-83.8)   Electrolyte\end{tabular}} & Density ${\rho}$ (kg/m$^{3}$)                                           & 2,690          & \multirow{6}{*}{{}\cite{Cornwell1971TheSalts,Masset2007ThermalElectrolytes}{}}       \\
                                                                                                           & Volumetric thermal expansion coefficient ${\beta_{\mathrm{Th}}}$ (1/K) & 3.17x10$^{-4}$      &                                    \\
                                                                                                           & Kinematic viscosity \textit{v} (m$^{2}$/s)                                                 & 7.77x10$^{-7}$     &                                    \\
                                                                                                           & Heat capacity ${c_{\mathrm{p}}}$ (J/kgK)                                                  & 1,220          &                                    \\
                                                                                                           & Thermal conductivity ${\lambda}$ (W/mK)                          & 0.248          &                                    \\
                                                                                                           & Electrical conductivity ${\sigma}$ (S/m)           & 288            &                                    \\ \hline
\multirow{8}{*}{Pb(Li) Cathode}                                                                            & Density ${\rho}$ (kg/m$^{3}$)                                           & 8,960          & \multirow{8}{*}{{}\cite{Herreman2020SolutalBatteries,Khairulin2017VolumetricMelts,Kondo2016EvaluationAlloys,Agency2015HandbookTechnologies}{}} \\
                                                                                                           & Volumetric thermal expansion coefficient ${\beta_{\mathrm{Th}}}$ (1/K) & 1.37x10$^{-4}$     &                                    \\
                                                                                                           & Solutal thermal expansion coefficient ${\beta_{\mathrm{S}}}$ (m3/kg)       & 1.31x10$^{-3}$     &                                    \\
                                                                                                           & Diffusion coefficient \textit{D} (m$^{2}$/s)                                             & 8x10$^{-9}$         &                                    \\
                                                                                                           & Kinematic viscosity \textit{v} (m$^{2}$/s)                                                & 1.61x10$^{-7}$      &                                    \\
                                                                                                           & Heat capacity ${c_{\mathrm{p}}}$ (J/kgK)                                                   & 147            &                                    \\
                                                                                                           & Thermal conductivity ${\lambda}$ (W/mK)                          & 14.9           &                                    \\
                                                                                                           & Electrical conductivity ${\sigma}$ (S/m)                         & 1,380,000      &                                    \\ \hline \hline
\end{tabular}
\end{table}
\vspace{-0.5pc}
The two capacities considered in this work are 241 Ah and 481 Ah. These capacities correspond to areal capacities, the capacity per unit area, of 3.75 Ah/cm$^{2}$ and 7.5 Ah/cm$^{2}$. This is similar to those tested by Zhou \textit{et al.}, in which the final battery had a capacity of 400 Ah and is the largest capacity cell reported to date \cite{Zhou2022IncreasingBattery}. These capacities are ideal for long-duration energy storage and will be required in grid-scale installations of LMBs \cite{Zhou2022IncreasingBattery}. The corresponding layer heights are provided in Table 3, and the dimensions can be matched to Fig 2.
\begin{table}[ht]
\caption{Layer heights for the numerical geometries modelled in this work. Dimensions are in mm}
\centering
\begin{tabular}{lcc}
\hline
\hline
Variable cell dimensions in Fig 2. & 241 Ah & 481 Ah \\ \hline
${h_\mathrm{Li}}$                              & 20     & 40     \\
${h_\mathrm{El}}$                              & 10     & 10     \\
${h_\mathrm{Pb(Li)}}$                        & 10     & 20     \\
${h_\mathrm{C}=0.5h_\mathrm{Pb(Li)}}$                & 5      & 10     \\ \hline \hline
\end{tabular}
\end{table}
\subsection{Numerical model}
\label{Numerical model}
The numerical model developed by Keogh \textit{et al.} was used in this work \cite{Keogh2021ModellingBatteries,Keogh2023ModellingBatteries}. It has been validated against experimental results for convection and electro-vortex flow in a single layer of Pb-Bi \cite{Ashour2018CompetingBatteries,Keogh2021ModellingBatteries}, three-layer results from a validated pseudo-spectral solver \cite{Personnettaz2018ThermallyBatteries,Kollner2017ThermalModel,Keogh2023ModellingBatteries}, and the charge/discharge curves of a Li${||}$Bi battery \cite{Personnettaz2019MassBatteries,Keogh2023ModellingBatteries}. The numerical model is implemented in OpenFOAM and uses a multi-region approach where domains are coupled at shared interfaces. It uses the quasi-static approximation, which is only valid if the magnetic Reynolds number is small, typically ${R_\mathrm{m}<<1}$. The magnetic Reynolds number is given by
\begin{equation}
R_\mathrm{m}=\sigma\mu_\mathrm{0}ul
\end{equation}
in which ${\sigma}$ is the electrical conductivity, ${\mu_\mathrm{0}}$ is the magnetic permeability of the medium, taken as the vacuum permeability in this case, \textit{u} is a characteristic velocity, and \textit{l} is a characteristic length scale. This dimensionless number estimates the importance of the advection and diffusion of the magnetic field. If the magnetic Reynolds number is small, then the magnetic field is well determined by the boundary conditions and any perturbations to the field due to the fluid velocity are insignificant. Using an ${R_\mathrm{m}}$ value of 0.01, taking the electrical conductivity of lead from Table 2, and the height of the lead layer in the largest case, \textit{l} = 20 mm, it was found that so long as the velocity was less than 268 mm/s this limit was met. A typical maximum velocity found in the computations was 90 mm/s, so these computations fall in the low ${R_\mathrm{m}}$ limit.
\subsubsection{Static and induced electric and magnetic fields}
\label{Static and induced electric and magnetic fields}
The formulation used in this is the same as described by Davidson \cite{Davidson2010DynamicsNumbers}. The quasi-static approximation assumes that there are two components to the electric field. A stationary component from the applied field that remains fixed throughout the computations, and an induced component from the fluid velocity without a time derivative. The magnetic field remains fixed, and this field has no induced components. Thus, the electric potential and current density are considered as summations of the static and induced components
\begin{equation}
\phi=\phi_\mathrm{a}+\phi_\mathrm{i},
\end{equation}
\begin{equation}
\textbf{\textit{J}}=\textbf{\textit{J}}_\mathrm{a}+\textbf{\textit{J}}_\mathrm{i},
\end{equation}
in which ${\phi}$ is the electric potential, \textbf{\textit{J}}  is the current flux vector, and the subscripts a and i indicate whether the field is applied or induced. The Laplace equation for electric potential is used to calculate the applied potential
\begin{equation}
\nabla\cdot\sigma\nabla\phi_\mathrm{a}=0,
\end{equation}
with boundary conditions
\begin{equation}
\phi_\mathrm{a}=0,
\end{equation}
\begin{equation}
\nabla\phi_\mathrm{a}=-\frac{\textbf{\textit{J}}_\mathrm{a}}{\sigma}\hat{\textbf{z}},
\end{equation}
applied at opposing current collectors. From the electric potential, the current density is determined using
\begin{equation}
\textbf{\textit{J}}_\mathrm{a}=-\sigma\nabla\phi_\mathrm{a}.
\end{equation}
The magnetic field is then computed with the magnetic vector potential formulation using a coulomb gauge which ensures the field is divergence-free
\begin{equation}
\nabla^{2}\textbf{\textit{A}}=-\mu_\mathrm{0}\textbf{\textit{J}}_\mathrm{a},
\end{equation}
in which \textbf{\textit{A}} is the magnetic vector potential. The magnetic flux vector is obtained from the curl of the magnetic vector potential such that 
\begin{equation}
\textbf{\textit{B}}=\nabla\times\textbf{\textit{A}},
\end{equation}
which gives the corresponding magnetic field to the current inside the cell and embedded coil. To simplify the computations, the current density inside the coil is assumed to be homogeneous and solely in the azimuthal direction. These fields make up the static component of the electric field. The induced components of the electric fields are calculated using
\begin{equation}
\nabla\cdot\sigma\nabla\phi_i=\nabla\cdot(\sigma(\textit{\textbf{u}} \times \textit{\textbf{B}})),
\end{equation}
\begin{equation}
\textbf{\textit{J}}_\mathrm{i}=\sigma(-\nabla\phi_\mathrm{i}+\textbf{\textit{u}}\times\textbf{\textit{B}}),
\end{equation}
in which \textbf{\textit{u}} is the fluid velocity vector. A reference point is used to set a zero-reference potential, and the induced electric potential is then solely calculated from the fluid velocity and the background magnetic field found using Equation (9). The potential fields are coupled between regions using a Dirichlet-Neumann boundary condition such that the potential at the interface is given by
\begin{equation}
\phi_\mathrm{i}=\frac{\frac{\phi_\mathrm{icc}\sigma_\mathrm{icc}}{\delta_\mathrm{icc}}+\frac{\phi_\mathrm{jcc}\sigma_\mathrm{jcc}}{\delta_\mathrm{jcc}}}{\frac{\sigma_\mathrm{icc}}{\delta_\mathrm{icc}}+\frac{\sigma_\mathrm{jcc}}{\delta_\mathrm{jcc}}},
\end{equation}
in which ${\phi_\mathrm{i}}$ is the potential at the interface, ${\phi_\mathrm{cc}}$ is the potential at the cell centre adjacent to the interface, ${\sigma_\mathrm{cc}}$ is the electrical conductivity at the cell centre adjacent to the interface, ${\delta_\mathrm{cc}}$ is the distance from the interface to the cell centre, and the subscript i and j indicate the fluid on either side of the interface. 
\subsubsection{Mass transfer}
\label{Mass transfer}
The approach used by in this numerical model to calculate mass transfer in the battery is the same as has been used extensively in the literature \cite{Personnettaz2019MassBatteries,Ashour2018Convection-diffusionBatteries,Weber2020NumericalFlow,Herreman2020SolutalBatteries, Herreman2021EfficientBatteries, Personnettaz2022LayerBatteries}. A scalar transport equation,
\begin{equation}
\frac{\partial\psi}{\partial\textit{t}}+\nabla\cdot(\textbf{\textit{u}}\psi)-\nabla^{2}(\textit{D}\psi)=0,
\end{equation}
in which ${\psi}$ is the mass concentration of lithium, captures the essential characteristics of the convection and diffusion of lithium into lead as anodic atoms are transported from the cathode-electrolyte interface into the cathode. The key assumption when using this approach is that the height of the cathode changes by a small enough height during the computations that it can be neglected. This limits the timeframe of the computations to changes in molar fractions of no more than \(\Delta x_\mathrm{Li}=0.1\) \cite{Weber2022CellBatteries}. If this condition is satisfied then the discharge voltages of the battery can be captured with errors as low as 0.5\(\%\) \cite{Personnettaz2019MassBatteries}.
Since there are no changes in height, the electric field does not change during the computations. The mass flux, or gradient of the concentration field, can then be calculated using the current flux from the applied electric field. Faraday's law is applied at the cathode-electrolyte interface
\begin{equation}
\nabla\psi=-\frac{J_\mathrm{a}M_\mathrm{Li}}{FD}\hat{\textbf{z}},
\end{equation}
in which F is Faraday's constant, \(M_\mathrm{Li}\) is the molar mass of lithium and \(\hat{\textbf{z}}\) is the normal direction vector.
\subsubsection{Heat Transfer}
\label{Heat transfer}
In this model, the only source of heat is the Joule heating from the discharge current. This consideration is consistent with multiple previous studies of LMBs \cite{Kollner2017ThermalModel, Zhou2022Multi-fieldPerformance,Personnettaz2022LayerBatteries, Shen2016ThermalBattery, Keogh2023ModellingBatteries}. The energy equation is used to this end,
\begin{equation}
\frac{\partial T}{\partial t}+(\textbf{\textit{u}}\cdot\nabla)T=\frac{\lambda}{\rho_\mathrm{0}c_\mathrm{p}}\nabla^{2}T+\frac{J^{2}}{\sigma\rho_\mathrm{0}c_\mathrm{p}},
\end{equation}
in which ${\lambda}$ is the thermal conductivity, ${\rho_\mathrm{0}}$ is the reference density and ${c_\mathrm{p}}$ is the specific heat capacity. This equation is coupled between regions with a Dirichlet-Neumann boundary condition such that the temperature at the interface is given by
\begin{equation}
T_\mathrm{i}=\frac{\frac{T_\mathrm{icc}\lambda_\mathrm{icc}}{\delta_\mathrm{icc}}+\frac{T_\mathrm{jcc}\lambda_\mathrm{jcc}}{\delta_\mathrm{jcc}}}{\frac{\lambda_\mathrm{icc}}{\delta_\mathrm{icc}}+\frac{\lambda_\mathrm{jcc}}{\delta_\mathrm{jcc}}},
\end{equation}
in which \(T_\mathrm{i}\) is the temperature at the interface, \(T_\mathrm{cc}\) is the temperature at the cell centre adjacent to the interface and \(\lambda_\mathrm{cc}\) is the thermal conductivity at the cell centre adjacent to the interface. To solve this equation, it was assumed that the sidewalls of the LMB are thermally non-conducting, an assumption justified by the low thermal conductivity of Al\(_\mathrm{2}\)O\(_\mathrm{3}\) \cite{White1983Thermal3}, and that the top and bottom boundary maintain a constant temperature equivalent to the operating temperature of the battery, ${T}$ = 500\textdegree C. 
\subsubsection{Fluid dynamics}
\label{Fluid Dynamics}
The incompressible formulation of the Navier-Stokes and continuity equations are used to model the fluid flow in the layers,
\begin{equation}
\frac{\partial\textbf{\textit{u}}}{\partial t}+\textbf{\textit{u}}\cdot\nabla\textbf{\textit{u}}-\textit{v}\nabla^{2}\textbf{\textit{u}}=\frac{1}{\rho_\mathrm{0}}\nabla(\rho_\mathrm{0}\textit{\textbf{uz}}-p)-\textit{\textbf{g}}(\beta_\mathrm{Th}(T-T_\mathrm{0})+\beta_\mathrm{s}(\psi-\psi_\mathrm{0}))+\frac{\textbf{\textit{f}}_\mathrm{L}}{\rho_\mathrm{0}},
\end{equation}
\begin{equation}
\nabla\cdot\textbf{\textit{u}}=0,
\end{equation}
in which ${\textit{p}}$ is pressure, ${\textbf{\textit{f}}_\mathrm{L}}$ is the Lorentz body force, ${\textbf{\textit{z}}}$ is the coordinate vector and ${\textbf{\textit{g}}}$ is the gravitational acceleration vector. The Lorentz body force is given by
\begin{equation}
\textbf{\textit{f}}_\mathrm{L}=\textbf{\textit{J}}\times\textbf{\textit{B}}.
\end{equation}
These equations are solved by assuming that the interfaces between the layers are impermeable and unmoving. This assumption appears to be valid as in the experiment there was no visible deformation or vibration of the interface as shown by supplementary video 1, which is a video of the top surface with current applied. 
\par
The state of each layer is solved individually on its own mesh, with momentum transferred between the layers through a boundary condition at the interface that accounts for the viscous coupling between layers. This boundary condition was derived and implemented by Weber \textit{et al.} \cite{Weber2020NumericalFlow} and Personnettaz \textit{et al.} \cite{Personnettaz2022LayerBatteries}. It can be found in \cite{Personnettaz2022LayerBatteries} and \cite{Keogh2023ModellingBatteries} but, for completeness, it is given as follows; the coupled velocity at an interface is given by
\begin{equation}
\textbf{\textit{u}}_\mathrm{f1}=\textbf{\textit{u}}_\mathrm{f2},
\end{equation}
\begin{equation}
\textbf{\textit{u}}\cdot\textbf{\textit{n}}=0,
\end{equation}
\begin{equation}
\mu_\mathrm{1}\nabla\textbf{\textit{u}}_\mathrm{t1}\cdot\textbf{\textit{n}}_\mathrm{1}=\mu_\mathrm{2}\nabla\textbf{\textit{u}}_\mathrm{t2}\cdot\textbf{\textit{n}}_\mathrm{2},
\end{equation}
in which ${u_\mathrm{f}}$ is the velocity at the coupled interface, ${u_\mathrm{t}}$ is the tangential velocity at the cell centre adjacent to the interface, ${u_\mathrm{tf}}$ is the tangential velocity at the face, \textbf{\textit{n}} is the normal vector and ${\mu}$ is the dynamic viscosity. Combining Equation (21) and (22) gives
\begin{equation}
\mu_\mathrm{1}\frac{\textbf{\textit{u}}_\mathrm{t1}-\textbf{\textit{u}}_\mathrm{tf}}{\delta_\mathrm{2}}=\mu_\mathrm{2}\frac{\textbf{\textit{u}}_\mathrm{t2}-\textbf{\textit{u}}_\mathrm{tf}}{\delta_\mathrm{2}}.
\end{equation}
Rearranging for ${\textbf{\textit{u}}_\mathrm{tf}}$ gives
\begin{equation}
\textbf{\textit{u}}_\mathrm{tf}=w\cdot\textbf{\textit{u}}_\mathrm{t1}+(1-w)\cdot\textbf{\textit{u}}_\mathrm{t2},
\end{equation}
in which w is the weighting factor given by
\begin{equation}
w=\frac{\delta_\mathrm{2}\cdot\mu_\mathrm{1}}{\delta_\mathrm{1}\mu_\mathrm{2}+\delta_\mathrm{2}\mu_\mathrm{1}}.
\end{equation}
Since a tangential velocity can be defined as
\begin{equation}
\textbf{\textit{u}}_\mathrm{t}=\textbf{\textit{u}}\cdot(\textbf{\textit{I}}-\textbf{\textit{nn}}),
\end{equation}
the boundary condition for the velocity at the interface can then be written as
\begin{equation}
\textbf{\textit{u}}_\mathrm{tf}=w\cdot\textbf{\textit{u}}_\mathrm{1}\cdot(\textbf{\textit{I}}-\textbf{\textit{nn}})+(1-w)\cdot\textbf{\textit{u}}_\mathrm{2}\cdot(\textbf{\textit{I}}-\textbf{\textit{nn}})
\\
=(\textbf{\textit{I}}-\textbf{\textit{nn}})\cdot(w\cdot\textbf{\textit{u}}_\mathrm{1}+(1-w)\cdot\textbf{\textit{u}}_\mathrm{2}),
\end{equation}
in which \textbf{\textit{I}} is the identity matrix. For all other surfaces, a no-slip boundary condition is used.
\subsubsection{Explicitly estimating cell potential}
\label{Explicitly estimating cell potential}
Estimation of the electrochemical performance of the cell is quantitatively determined by explicitly estimating the cell potential from the interfacial concentration. This method has been successfully validated against the discharge curves of LMBs in \cite{Personnettaz2019MassBatteries,Weber2022CellBatteries, Keogh2023ModellingBatteries}. Since the electrochemical potential of batteries is an interfacial phenomenon, the first step is to determine the average interfacial concentration which can be defined as
\begin{equation}
\langle \psi \rangle _{S}=\frac{1}{\pi R_\mathrm{C} ^{2}}\int_{S} \psi |_{z=H_\mathrm{E}}dS,
\end{equation}
in which ${\langle \psi \rangle _{S}}$ is the instantaneous average interfacial concentration, ${R_\mathrm{C}}$ is the radius of the cathode, and ${H_\mathrm{E}}$ indicates that the integral is taken at the cathode-electrolyte interface. The next step is to determine the molar fraction of lithium at the interface where
\begin{equation}
x_\mathrm{Li}=\frac{n_\mathrm{Li}}{n_\mathrm{Li}+n_\mathrm{Pb}},
\end{equation}
is the molar fraction of lithium and \textit{n} is the number of moles. This requires the computation of the number of moles in each cell volume which is unknown. However, the concentration at each cell volume is computed so substituting for the concentration using
\begin{equation}
n_\mathrm{Li}=\frac{\psi v_\mathrm{D}}{M_\mathrm{Li}},
\end{equation}
in which ${\textit{v}_\mathrm{D}}$ is the cell volume, allows the mass fraction to be found using
\begin{equation}
x_\mathrm{Li}=\frac{\langle \psi \rangle _{S} M_\mathrm{Pb}}{\langle \psi \rangle _{S} M_\mathrm{Pb}+ \langle \psi \rangle _{SPb} M_\mathrm{Li}},
\end{equation}
which can be computed since an operating principle of the numerical model is that the lead concentration remains the same throughout the volume in time. The open circuit voltage (OCV) is then interpolated from the emf data of Gasior and Moser \cite{Gasior2001ThermodynamicMethod}, which was determined at molar fractions of lithium in lead, ${x_\mathrm{Li}}$, between 0.025 and 0.965. The nominal voltage of the battery can be determined by taking the time integral of the surface concentration data using
\begin{equation}
\overline{\langle \psi \rangle} _{S}=\frac{1}{t}\int_{t} \langle \psi \rangle _{S} dt,
\end{equation}
and then using this value to interpolate for the OCV. Another key parameter is the volume averaged concentration which can be determined by taking the volume integral of the concentration field
\begin{equation}
\langle \psi \rangle _{V}=\frac{1}{\pi R_\mathrm{C} ^{2} H}\int_{V} \psi |_{z=H_\mathrm{E}}dV.
\end{equation}
The volume averaged concentration allows for the mass transport overpotential to be calculated by interpolating for the surface and volumetric OCV's and taking the difference. The divergence of the surface averaged concentration from the volume averaged concentration is used to test for convergence of the numerical model and this is calculated using
\begin{equation}
\nabla\cdot\psi=\langle \psi \rangle _{S}-\langle \psi \rangle _{V}.
\end{equation}
So that the additional parasitic losses of the coil can be quantified and used to estimate the true performance of the modified design, the ohmic overpotentials of the coil are calculated using
\begin{equation}
\phi_\mathrm{coil}=\frac{J_\mathrm{coil}L_\mathrm{coil}}{\sigma_\mathrm{c}},
\end{equation}
in which ${L_\mathrm{coil}}$ is the length of the coil, which is taken as 76.8\ mm and can be estimated from Fig. 2.
\subsubsection{Convergence}
\label{Convergence}
Convergence of the numerical model was determined by comparing the divergence of the concentration at the cathode-electrolyte interface from the volumetric concentration for 160 seconds of computation time for three different mesh sizes in the model with a 10 mm cathode layer height. Table 4 shows that a mesh with quadrilaterals with a side length 1.25${\times{10}^{-4}}$ m is converged for interfacial concentration. This mesh was used to analyse the performance of the battery. The non-uniform mesh has better resolution, particularly around the current collector, and so this discretisation was used to analyse the dynamics of the flow. For each case, the mesh was purely quadrilateral and structured with five additional layers applied around the walls to resolve the boundary layer.
\par
Table 5 shows the mesh sizes used for the additional cases. Since the numerical domain was complex, a variable time step was used to ensure efficient use of computational resources. So that the computations remained fully resolved in time, two constraints were set on the timestep size; the courant number was limited to a value of 0.5, and the maximum timestep size was limited to ${t=2.5\times{10}^{-4}}$ s. Due to these constraints on the step size, typical courant numbers were found to be in the range of 0.2 - 0.4. A second-order accurate discretisation scheme was used with linear interpolation for all terms and the backwards Euler time scheme and a convergence criterion of 1${\times{10}^{-8}}$. The concentration fields were solved with a fourth-order cubic scheme. The PIMPLE method, which is a combination of the Pressure Implicit Splitting of Operators (PISO) and Semi-Implicit Method for Pressure Linked Equations (SIMPLE) methods, was used to solve the pressure equation and, owing to the explicit coupling of the velocity fields, at every time step the equations in each region were solved three times with two pressure corrector steps during each solution.
\begin{table}[ht]
\caption{Convergence study for the numerical modelling. The mesh in the electrolyte and anode were similarly refined during each study.}
\label{tab:convergence}
\centering
\begin{tabular}{ p{3.7cm} p{3.7cm} p{3.7cm} p{3.7cm} }
 \hline
 \hline
 \textbf{Cell side length (m)}& \textbf{Number of cells in the cathode layer}& \textbf{${\nabla\cdot\psi}$ (kg/m$^{3}$)}& \textbf{Percent difference (${\%}$)} \\
 \hline
 ${1.5\times10^{-4}}$ & ${2.7\times10^{4}}$& 13.132& -\\
 ${1.2\times10^{-4}}$ & ${3.7\times10^{4}}$& 13.482& 2.63\\
 ${8.5\times10^{-5}}$ - ${1.35\times10^{-4}}$ & ${4.6\times10^{4}}$& 13.495& 0.1\\
 \hline
 \hline
\end{tabular}
\end{table}
\vspace{-1pc}
\par
\begin{table}[ht]
\caption{Mesh size used for the computation of each case.}
\label{tab:mesh}
\centering
\begin{tabular}{ p{5cm} p{3cm} }
 \hline
 \hline
 \textbf{Case}& \textbf{Cell side length (m)} \\
 \hline
 5,000 A/m$^{2}$ & ${1.25\times10^{-4}}$ \\
 3,000 A/m$^{2}$ & ${1.25\times10^{-4}}$ \\
 10,000 A/m$^{2}$ & ${7.5\times10^{-5}}$\\
 5,000 A/m$^{2}$ Large Capacity & ${1.25\times10^{-4}}$ \\
 Axisymmetric experiment & ${6.5\times10^{-5}}$\\
 \hline
 \hline
\end{tabular}
\end{table}
\vspace{-1pc}
\par
\subsection{Experimental Method}
\label{Experimental Method}
The design was tested using a lab-scale prototype and PIV with the setup illustrated in Fig. 3. The coil in the experiment was machined from copper and mounted in an acrylic fitting with the copper current supply. Both the copper current supply and coil were isolated from the working fluid using an acrylic vinyl coating. The diameter of both the coil and current supply was ${\textit{D}=3.2}$ mm, and the thickness of the acrylic vinyl coating was 0.2 mm. The coil and current supply were then connected in series such that the current delivered to the current supply travelled through the coil first, generating a vertical magnetic field. The acrylic fitting was mounted in a copper plate using an interference fit. The copper plate served as the current collector, and the current was forced to enter the working fluid owing to the insulation provided by the acrylic fitting and acrylic vinyl coating before exiting through the copper plate. The vessel, which was made from cast Plexiglas with an internal diameter of ${D=44}$ mm, was mounted directly on top of the copper current collector.
\par
The working fluid used was the alloy Galinstan at a composition of Ga\(^{67}\)In\(^{20.5}\)Sn\(^{12.5}\). The laboratory temperature during the experiments was determined as ${T=22.6}$\textdegree C at the time of the experiments. The material properties of Galinstan at this temperature are ${\sigma=3.27\times10^{6}}$ S/m, ${\rho=6440}$ kg/m$^{3}$, ${v=3.302\times10^{-7}}$ m$^{2}$/s \cite{Plevachuk2014ThermophysicalAlloy}. Upon exposure to the atmosphere, liquid metals form an oxide coating on the external surfaces which gives boundary conditions close to no-slip. To dissolve this oxide coating, a thin layer of 6.4${\%}$ HCl solution was added to the surface. This concentration was used as it enables the oxide formed on the surface of the liquid metal to be dissolved at a rate faster than the rate of reaction with any oxygen absorbed in the acid solution \cite{Xu2012EffectGallium-indium}. The material properties of HCl at this temperature are ${\rho=1045.14}$ kg/m$^{3}$ and ${v=1.016\times10^{-6}}$ m$^{2}$/s \cite{Nishikata1981ViscositiesSolutions}.
\begin{figure}[htp]
    \centering
    \includegraphics[width=10.9cm]{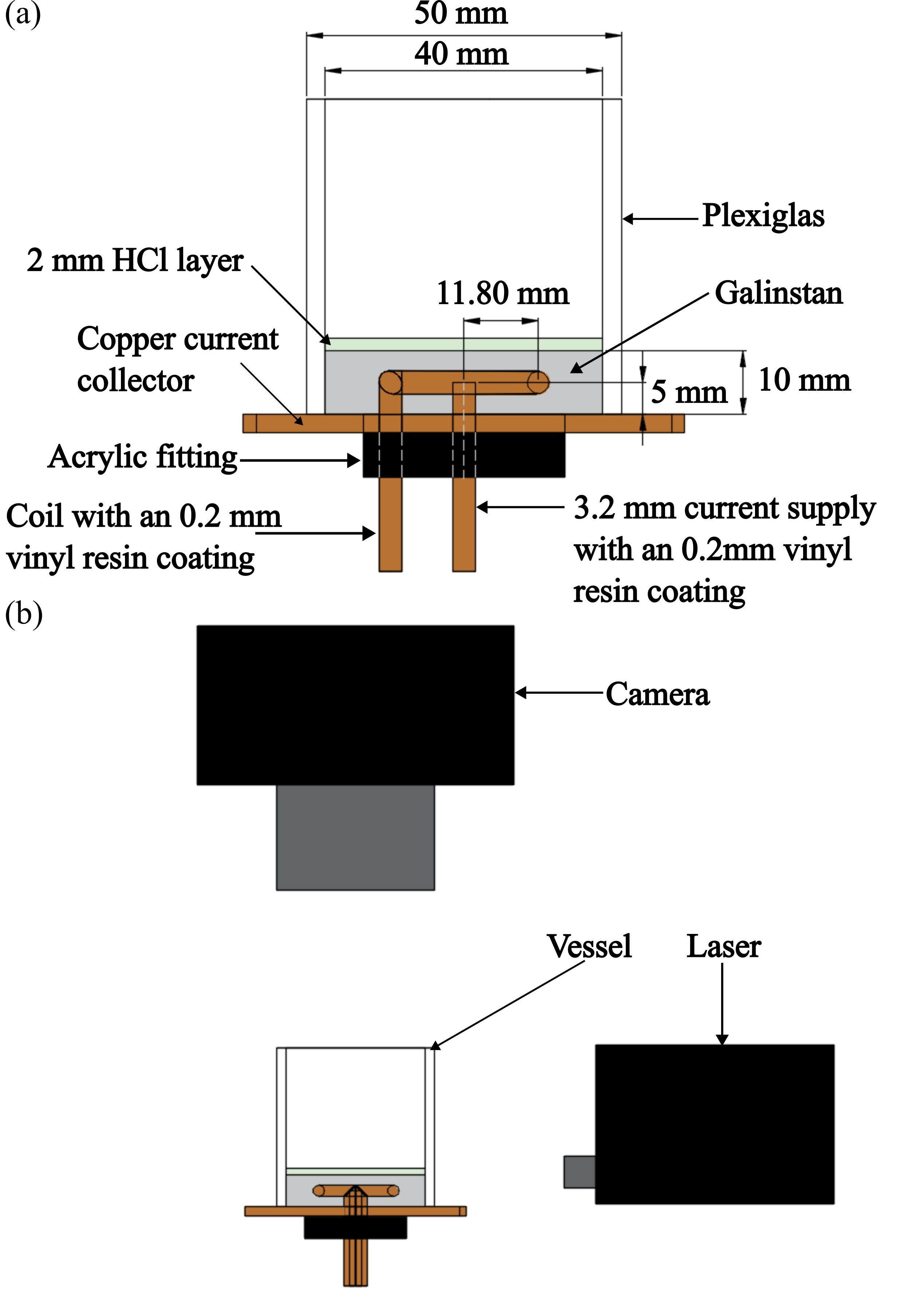}
    \caption{Experimental setup used for testing the design changes. (a) Vessel dimensions and layer heights used in the experiment (b) Simplified overview of the experimental setup used during measurements.}
    \label{fig:diagramExp}
\end{figure}
To take the PIV measurements the HCl was seeded with silver-coated glass microspheres with a diameter of 12 - 15 ${\mathrm{\mu m}}$ and a density of 1.00\ g/cm$^{3}$. The silver-coated glass microspheres are particularly well suited to being the seeding particle in HCl as silver will not react with low concentrations of HCl. A laser sheet 1 mm in width was directed through the HCl layer allowing the silver-coated particles to scatter light during the experiment. A Nikon D850 was mounted perpendicular to the layer, and footage of the fluid was taken at a frame rate of 120 frames per second, fps, with a shutter speed of 1/240 s, and an fstop of 2.8. The footage from the D850 was corrected for both tangential and radial distortion using 450 images of a checkerboard pattern taken before each run. These images were also used to determine the pixel-to-distance ratio. The height of the checkerboard pattern was ${h=11}$ mm from the top surface of the copper current collector which means that the camera was focused at ${h=11}$ mm, or the mid-height of the HCl layer, during the experiment. After correction, the images were analysed in PIVlab \cite{Thielicke2021ParticlePIVlab} using interrogation windows of 128x128, 64x64, 48x48 and 32x32 pixels with 50${\%}$ overlap. Linear correlation was used, and the final pass was repeated until the quality slope of the correlation coefficient was 0.025. After the frames were analysed, spurious data was removed using two filters: a standard deviation filter and a local median filter. Following this, any vectors that remained with a low correlation coefficient were removed, and interpolation was performed from the surrounding data. These filters only removed a limited number of vectors per frame, and these were often located around the perimeter of the Galinstan where the high surface tension of the fluid caused strong curvature, which created either reflections or a drop in luminosity depending on the location of the laser. The repeatability of the method was tested with three consecutive experimental runs, and the average velocity recorded in those runs is plotted in Fig. 4. For each repeat of the experiment, both the magnitude and temporal evolution of the velocity in the layer closely match. Temporal averages of the velocity after a statistically steady state was reached, taken as being from \textit{t} = 40 s onwards, show close agreement with the largest percent difference from the mean being 0.31${\%}$ as shown by Table 6. The data files for the PIV results are all attached as supplementary data.
\begin{table}[ht]
\caption{Repeatability tests of the experimental method. Repeatability was determined by comparison of the temporal average of repeat runs to the mean}
\label{tab:repeatability}
\centering
\begin{tabular}{ p{2cm} p{3.2cm} p{3.2cm} }
 \hline
 \hline
 \textbf{Repeat}& \textbf{Temporally averaged velocity (m/s)}& \textbf{Percent difference to mean (${\%}$)} \\
 \hline
 ${1}$ & ${14.76}$& 0.08\\
 ${2}$ & ${14.82}$& 0.31\\
 ${3}$ & ${14.74}$& 0.23\\
 \hline
 Mean & ${14.77}$& -\\
 \hline
 \hline
\end{tabular}
\end{table}
\vspace{-1pc}
\par
\begin{figure}[htp]
    \centering
    \includegraphics[width=9cm]{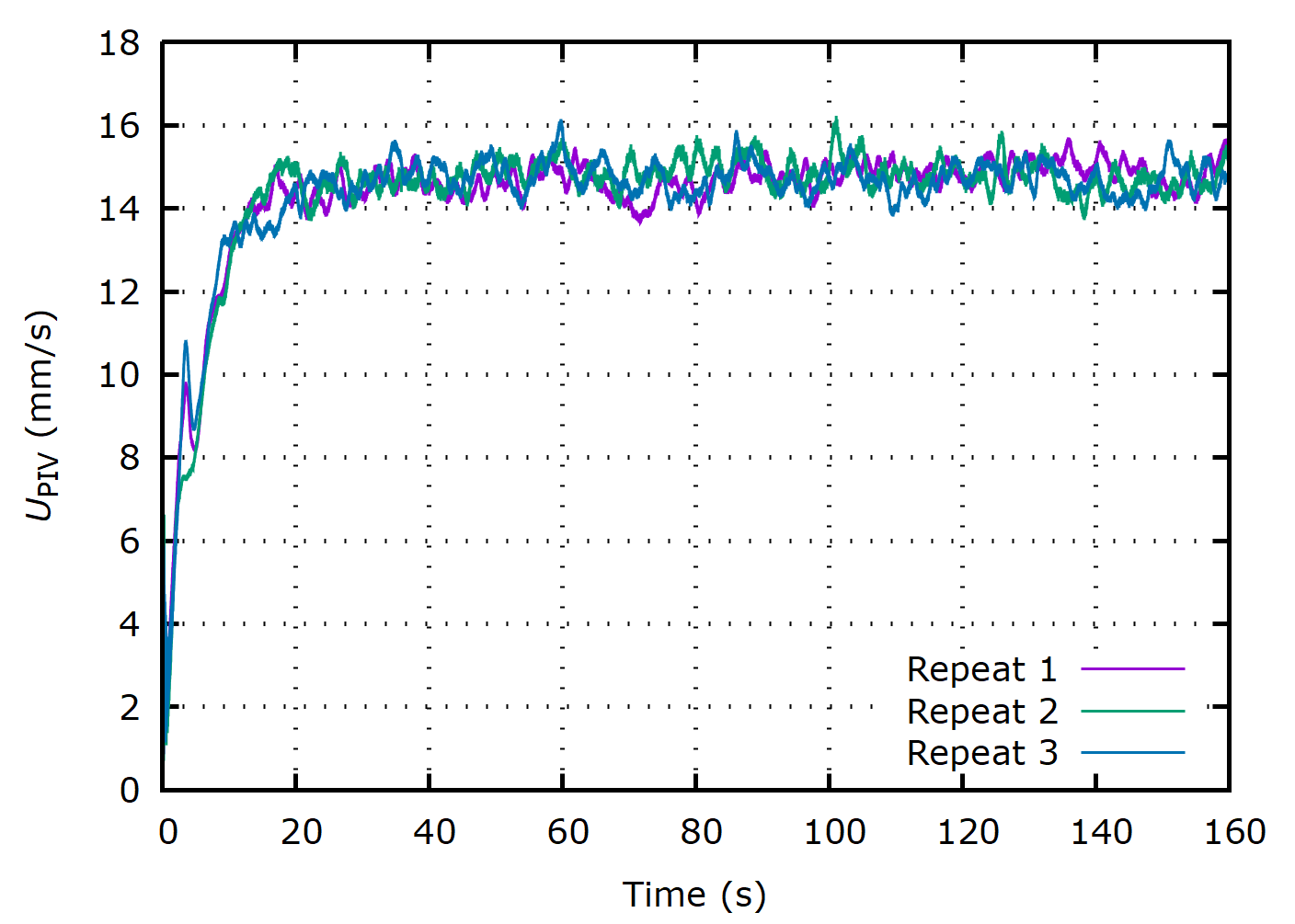}
    \caption{Comparison of $U_{\mathrm{PIV}}$ in repeatability tests}
    \label{fig:RepeatExp}
\end{figure}
\subsection{Key Parameters}
\label{Key Parameters}
All simulations performed in this work are axisymmetric. That is to say that the z-axis is taken to be the axis of rotational symmetry, and the velocity components are ${u(r,\theta,z)}$. This allows key global parameters to be defined as follows:
\begin{equation}
u=\sqrt{(\|u\|^{2})_V},
\end{equation}
in which u is the volume-averaged velocity magnitude. The poloidal velocity magnitude can be defined as:
\begin{equation}
u_\mathrm{pol}=\sqrt{\langle u_\mathrm{r}^{2}+u_\mathrm{z}^{2}\rangle_V},
\end{equation}
in which ${u_\mathrm{r}}$ and ${u_\mathrm{z}}$ are the velocity vectors parallel and perpendicular to the axisymmetric axis. Considering that the axisymmetric axis is vertical, the poloidal flow is important for mass transfer as it is responsible for the removal of ions from the cathode-electrolyte interface. The toroidal velocity magnitude is defined as:
\begin{equation}
u_\mathrm{tor}=\sqrt{\langle u_\mathrm{\theta}^{2}\rangle_V},
\end{equation}
in which ${u_\mathrm{\theta}}$ is the component of flow that rotates around the axisymmetric axis. Finally, the improvement of the old design over the new design is expressed using the formula for percent change
\begin{equation}
Percent\; change= \frac{|V_\mathrm{2}-V_\mathrm{1}|}{V1}\times100 \%,
\end{equation}
in which V is the value, and the subscripts 1 and 2 represent the old and new values, respectively.
\section{Results and Discussion}
\label{Results and Discussion}
The performance of the proposed design will be explored numerically under a number of different discharge conditions in section 4.1. This will be followed by an analysis of the fluid dynamics in the cathode in the computations in section 4.2. PIV measurements from a lab-scale prototype will be presented in section 4.3 before a discussion of how the design can be incorporated into battery modules in section 4.4.
\subsection{Performance}
The computations were performed at current densities of 0.3, 0.5, and 1 A/cm$^{2}$. For the 0.3 and 0.5 A/cm$^{2}$ cases the computations reached a change in molar fraction, ${\Delta x_\mathrm{Li}}$, of 0.1. The 1 A/cm$^{2}$ and the large capacity case were considerably more computationally expensive and were only computed until the behaviour of the interfacial concentration could be resolved. In all cases, the discharge current was determined using the interfacial area of the cathode-electrolyte interface. The current was then applied to the coil with a simplifying assumption that the current was homogeneous and completely azimuthal. This means that as the current in the cases changed, the magnetic field also changed proportionally. The reference cases were computed by assuming that there was no flow in the electrode. This assumption was shown by Personnettaz \textit{et al.} to capture the discharge behaviour of LMBs to within 0.5${\%}$ \cite{Personnettaz2019MassBatteries}, making it an accurate benchmark reference case.
\par
In the reference cases, once the computations began, the interfacial concentration diverged from the volumetric concentration sharply. Indeed in Fig. 5 (a) – (d), throughout the time computed, the gradient of the interfacial concentration was always higher than that of the volume-averaged concentration. That is to say that the surface-averaged interfacial concentration increased faster than the volumetric averaged concentration. In LMBs, the diffusion coefficient is small, in the order of ${\sim 1\times{10}^{-8}}$ m$^{2}$/s. This means that the diffusion of anodic atoms away from the cathode-electrolyte interface is slow. In fact, it is much slower than the mass flux of anodic atoms into the cathode, which is typically high in LMBs because of the low ohmic overpotentials and fast reaction kinetics. As the discharge current of the battery increased, the divergence of the interfacial concentration from the volumetric concentration also increased. This occurred because the flux of lithium atoms into the cathode increased, while the diffusion coefficient remained the same. The difference between the interfacial concentration and volume-averaged concentration for the 0.5 and 0.3 A/cm$^{2}$ cases after a ${\Delta x_\mathrm{Li}}$ of 0.1 is 124.83 and 82.97 kg/m$^{3}$, respectively. The time frame for the 1 A/cm$^{2}$ was considerably shorter, but comparing the divergence of the interfacial concentration from the bulk concentration between the three current densities at the same molar fraction, the concentration for 1 A/cm$^{2}$ was 86.3 kg/m$^{3}$ while it was 58.76 and 43.75 kg/m$^{3}$ for the 0.5 and 0.3 A/cm$^{2}$ cases, respectively. Similarly, the large-capacity cell also exhibits higher concentration polarisation between the interface and bulk of the fluid at the same molar fraction, although this is caused by an increase in discharge time. These results are reflected in the literature multiple times, where increases in mass transport overpotentials at higher discharge currents and areal capacities cause a reduction in voltage and coupled shortening of the discharge capacity below the theoretical capacity. Dai \textit{et al.} \cite{Dai2018CapacityBattery} found their test cell only utilised 50${\%}$ of its capacity at 1 A/cm$^{2}$ while Zhou \textit{et al.} \cite{Zhou2022IncreasingBattery} found that in their 400 Ah battery, the increase in capacity led to a maximum capacity utilisation of 84${\%}$ even at low discharge current densities.
\begin{figure}[htp]
    \centering
    \includegraphics[width=16cm]{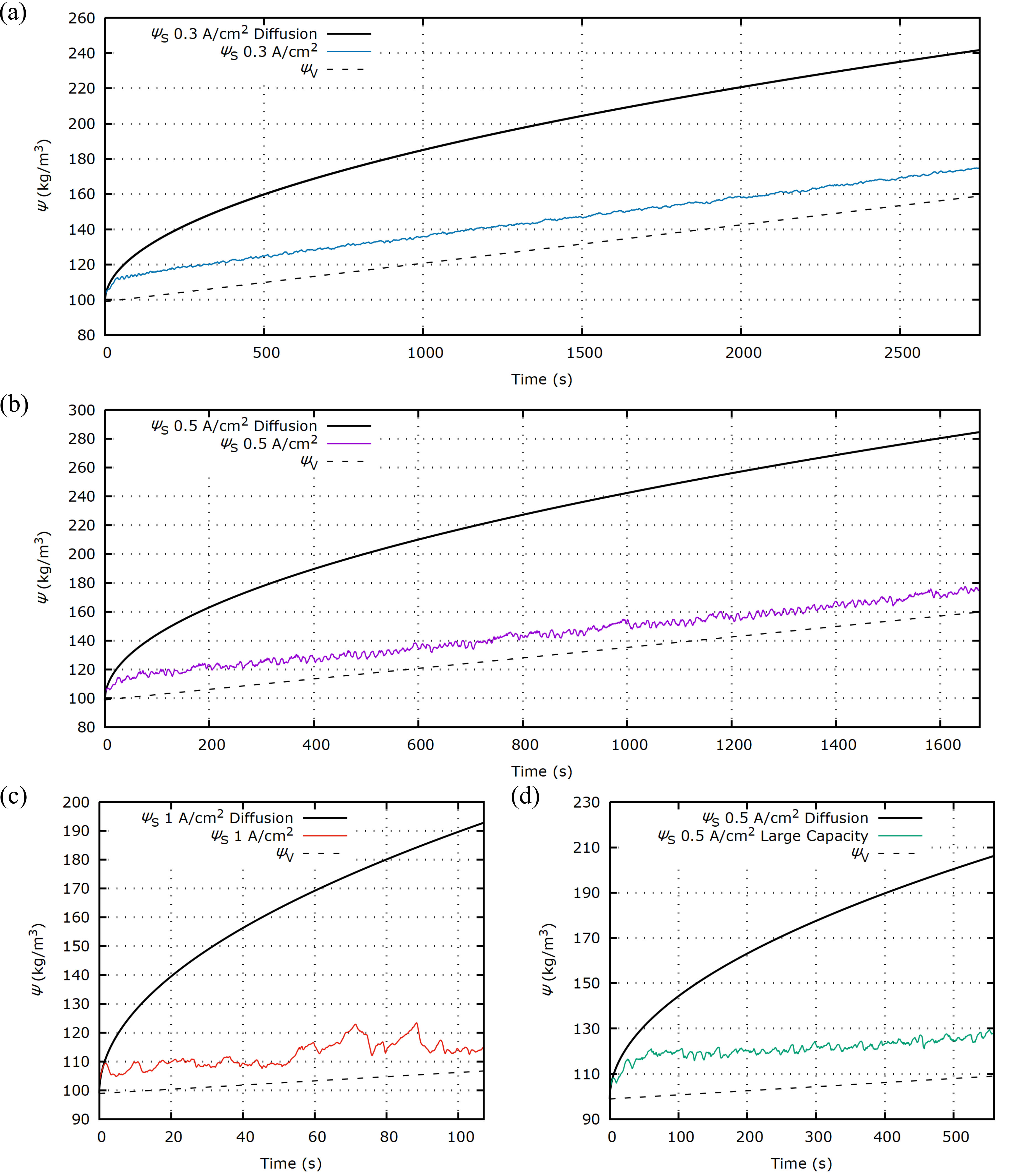}
    \caption{Instantaneous surface averaged concentration for all cases (a) 0.3 A/cm$^{2}$ (b) 0.5 A/cm$^{2}$ (c) 1 A/cm$^{2}$ (d) Large capacity case at 0.5 A/cm$^{2}$}
    \label{fig:performance}
\end{figure}
\par
These trends were not repeated in the modified LMB. In Fig. 5 (a) – (d), there was an initial transient period in which the concentration at the interface matched the diffusion cases. After this initial transient period, the flow induced by the coil mixed the cathode, homogenising the composition of the cathode and reducing the mass concentration of lithium at the cathode-electrolyte interface. The modified design had significantly lower concentration polarisations in the cathode; the divergence of the interfacial concentration decreased by 80.64${\%}$ for the 0.3 A/cm$^{2}$ cases and 87.64${\%}$ for the 0.5 A/cm$^{2}$ cases. Further, over the time frame of the computations, the concentration at the interface increased linearly as the volume-averaged concentration increased. The mixing was unaffected by the density stratification due to the Pb(Li) alloy, and the rate of mixing equalled the mass flux rate. The two cases presented here have an areal capacity of 3.75 Ah/cm$^{2}$ and 7.5 Ah/cm$^{2}$ which is a 7.5 hour and 15 hour discharge time at 0.5 A/cm$^{2}$. This indicates that if the flow remains unaffected, a LMB utilising this design could achieve complete capacity utilisation discharging at lengths suitable for long-duration storage applications such as renewable energy arbitrage \cite{Leadbetter2012SelectionElectricity}.
\par
The results were analysed to see how the average surface interfacial concentration was affected by the changing discharge current density. After a ${\Delta x_\mathrm{Li}}$ of 0.1, the difference between the interfacial and volume concentrations were 16.09 and 15.43 kg/m$^{3}$ for the 0.3 and 0.5 A/cm$^{2}$ cases, respectively. The surface averaged concentration at the final ${\Delta x_\mathrm{Li}}$ for the 1 A/cm$^{2}$ case was 7.31 kg/m$^{3}$ compared to 13.94 and 17.63 kg/m$^{3}$ for the 0.3 and 0.5 A/cm$^{2}$ cases at the same molar fraction. From this, it is clear that the modified design was unaffected by the increase in mass flux at higher current densities and, at 1 A/cm$^{2}$, there is an improvement in the surface averaged concentration. This behaviour was unexpected but occurred because the mixing rate increased faster than the mass flux. Taking the scaling laws presented by Herreman \textit{et al.} \cite{Herreman2021EfficientBatteries}, Davidson \textit{et al.} \cite{Davidson2022MagneticallyBattery}, and B\'enard \textit{et al.} \cite{Benard2022NumericalCylinders}, the velocity scales as either
\begin{equation}
\textbf{\textit{J}}\textbf{\textit{B}}^{\frac{2}{3}} \vee \textbf{\textit{J}}\textbf{\textit{B}}^{\frac{5}{9}}.
\end{equation}
Increasing the current density by a factor of 2 will also increase the magnetic field by a factor of 2 since the discharge current is channelled through the coil. Substituting these factors into the scaling laws, the velocity will increase by either 2.52 or 2.16, while doubling the discharge current will only lead to a doubling of the mass flow rate. This means that when discharging at higher current densities there is no additional mass-transport overpotential penalty. However, there are additional ohmic overpotentials that can be expected from both the coil and electrolyte layer.
\par
In Fig. 5 (d) the larger capacity cell displays slightly higher interfacial concentrations than the other cases, but the coil still mixed the cathode successfully. The longer time to velocity saturation in the cathode due to the larger volume caused the higher concentrations. Before the velocity was saturated, lithium atoms accumulated at the cathode-electrolyte interface until the mixing rate matched the mass flux rate.
\begin{table}[ht]
\caption{Divergence of the interfacial concentration from the volume-averaged concentration}
\label{tab:divergence}
\centering
\begin{tabular}{ p{3.2cm} p{3.2cm} p{2cm} }
 \hline
 \hline
 \textbf{Case}& \textbf{${\nabla\cdot\psi\ }$ at ${\Delta x_\mathrm{Li}=0.1}$ (kg/m$^{3}$) }& \textbf{Decrease (${\%}$)} \\
 \hline
 Diffusion 0.3 A/cm$^{2}$ & ${82.97}$& -\\
 0.3 A/cm$^{2}$ & ${16.09}$& 80.64\\
 Diffusion 0.5 A/cm$^{2}$ & ${124.83}$& -\\
 0.5 A/cm$^{2}$ & ${15.43}$& 87.64\\
 \hline
 \hline
\end{tabular}
\end{table}
\vspace{-1pc}
\par
The changes to the interfacial concentrations in the modified design have significant impacts on the electrochemical performance. Since mass transport overpotentials are coupled to the divergence of the concentration at the interface they are similarly reduced; there is a 78.31$\%$ reduction for the 0.3 A/cm$^{2}$ case and an 85.29$\%$ reduction for the 0.5  A/cm$^{2}$ case. This leads to an improvement in the voltage of the battery with the nominal OCV increasing by 6.32$\%$ and 9.55$\%$ for the 0.3 and 0.5 A/cm$^{2}$ cases, respectively. The difference between the references cases and the modified cases are much more significant at a ${\Delta x_\mathrm{Li}=0.1}$ (see Table 8), and if the full discharge cycle were modelled it can be expected that the differences would be much larger. The voltage increases should create much larger improvements in energy efficiency over a complete charge/discharge cycle, but this requires testing in a working LMB.
\newline
\newline\newline\newline
\par
\begin{table}[ht]
\caption{Electrochemical performance of the reference cases and the modified design}
\label{tab:performance}
\centering
\makebox[\textwidth]{%
\begin{tabular}{ p{1.5cm} p{1.5cm} p{1.5cm} p{1.5cm} p{1.5cm} p{1.5cm} p{1.5cm} p{1.5cm} p{1.5cm} }
 \hline
 \hline
 \textbf{Case}& \textbf{Nominal OCV (V)}& \textbf{Coil ohmic overpote\newline -ntial (V)} & \textbf{Net nominal OCV (V)} & \textbf{Percent increase (${\%}$)}& \textbf{OCV at ${\Delta x_\mathrm{Li}=0.1}$ } & \textbf{Percent increase (${\%}$)}& \textbf{Overpote\newline -ntials at ${\Delta x_\mathrm{Li}=0.1}$} & \textbf{Percent decrease (${\%}$)}\\
 \hline
 Diffusion 0.3 A/cm$^{2}$ \par & 0.4811 & - & 0.4811 & - & 0.4489 & - & 0.0590 & -\\
 0.3 A/cm$^{2}$ \par & 0.5208& 0.0093 & 0.5115 & 6.32 & 0.4859 & 8.24 & 0.0128 & 78.31\\
 Diffusion 0.5 A/cm$^{2}$ \par & 0.4618& - & 0.4618 & - & 0.4243 & - & 0.0836 & -\\
 0.5 A/cm$^{2}$ & 0.5214 & 0.0155 & 0.5059 & 9.55 & 0.4802 & 13.17 & 0.0123 & 85.29 \\
 \hline
 \hline
\end{tabular}}
\end{table}
\vspace{-1pc}
\par
\subsection{Flow Dynamics}
This section will analyse the results from the 0.5 A/cm$^{2}$ case in depth as it had the most distinct dynamics. The flow induced by the coil and current collector created improved operating conditions in the battery. As has been observed in past studies, the vertical magnetic field from the coil created a swirling flow which led to vigorous mixing. In typical swirling flows, the toroidal component of the velocity is an order of magnitude higher than the poloidal component \cite{Davidson1992SwirlingField}. Here, the poloidal component of the flow dominated the dynamics. Fig. 6 (a) quantifies the magnitudes of the toroidal and poloidal components observed, showing that the poloidal flow was typically twice the magnitude of the toroidal component.
Fig. 6 (b) and (c) depict the flow morphology. It consisted of a rapidly rotating jet with a velocity of 60 mm/s. The jet was dominated by a vertical action that created vortices that attached to the interface. These vortices mixed concentration gradients at the interface, enhancing mixing in the cathode. The interfacial vortices' impact is visible on the interfacial concentration, which oscillates in Fig. 5.
\begin{figure}[htp]
    \centering
    \includegraphics[width=16cm]{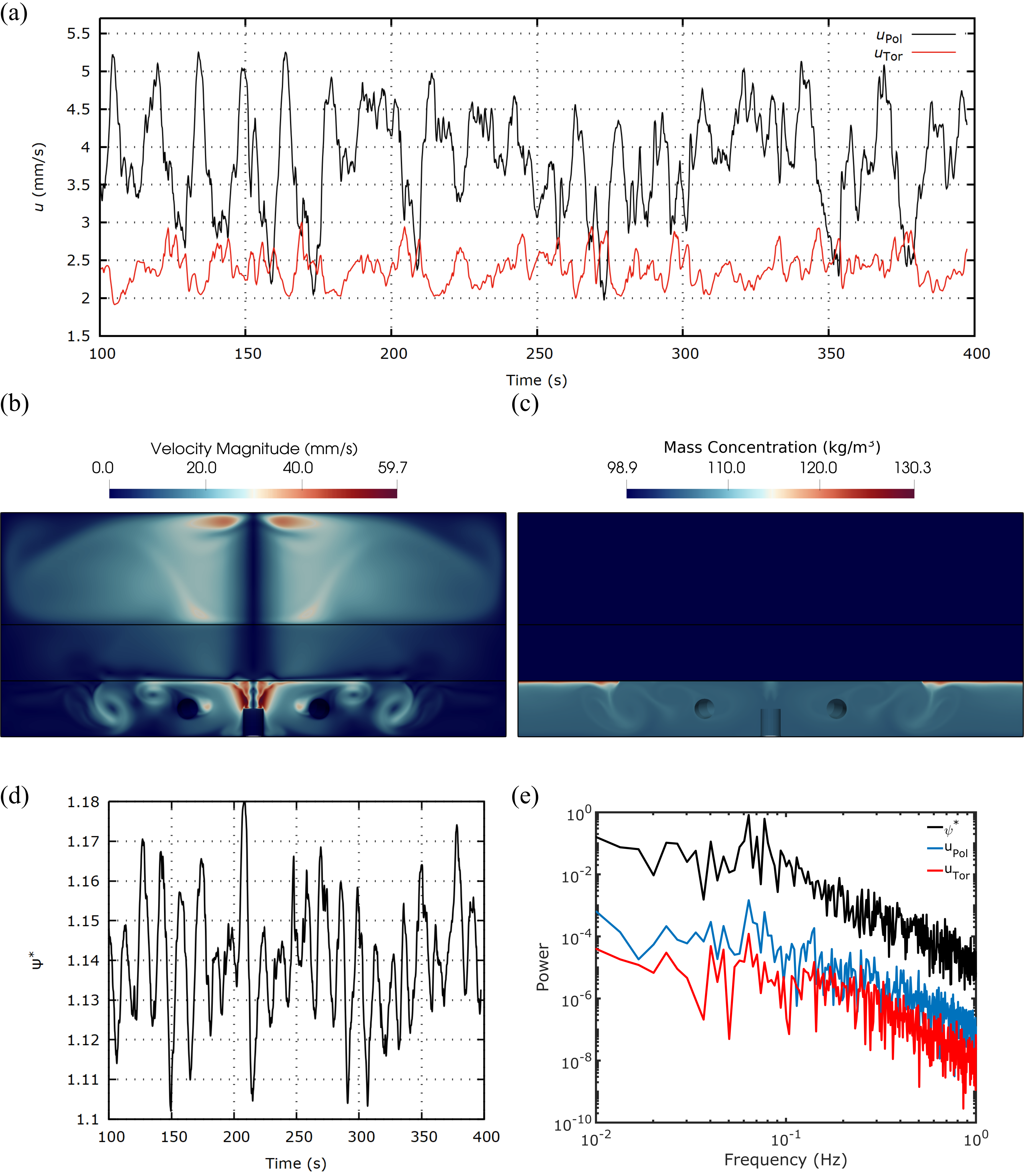}
    \caption{Flow dynamics at 0.5 A/cm$^{2}$ (a) Time evolution of the toroidal and poloidal velocity in the cathode layer (b) \& (c) Instantaneous snapshot of the velocity and concentration fields (d) Time evolution of the non-dimensionalised interfacial concentration (c) Characteristic frequencies of the vortex generation in the cathode layer}
    \label{fig:dynamics}
\end{figure}
\par
The vortices were created by the centrifugal instability commonly observed in Taylor-Couette flow. During discharge, the flow displayed cyclic behaviour. First, the fluid between the coil and the current collector accelerated. This acceleration was driven by the Lorentz force from the vertical magnetic field and horizontal current and was mainly toroidal. The resulting centrifugal force pushed the flow out radially, leading to an acceleration in the poloidal velocity. This is evident in Fig. 6 (a), which shows that the maxima of the toroidal flow were out of phase with the maxima of the poloidal flow. Taylor vortices have been found to be present in swirling flows in the battery in past works such as Davidson \textit{et al.} \cite{Davidson2022MagneticallyBattery}. The main difference between those found in this work and that of Davidson \textit{et al.} is that the geometrical configuration of the coil and current collector forced the vortices to travel along the interface.
\par
Like the Taylor vortices studied in the literature, the vortices observed here have temporal coherence. There is a clear periodicity in the velocity fluctuations in Fig. 6 (a), as well as the interfacial fluctuations in Fig. 5 (b). A Fourier analysis was conducted on the data to capture the peak frequencies of the flow. First, the interfacial concentration was non-dimensionalised by dividing by the volumetric averaged concentration, ${\psi^{*}=\frac{\langle \psi \rangle_\mathrm{S}}{\langle \psi \rangle_\mathrm{V}}}$, with the result shown in Fig. 6 (d). Then, a fast Fourier transform of the data was performed, and the result is plotted in Fig. 6 (e). There are two peak frequencies at 0.0637 and 0.0771 Hz. Two peak frequencies are found due to the small amount of data used to resolve the pattern, and a longer sampling time would likely see these two frequencies collapse into a signal peak. These frequencies represent the rate at which the interfacial vortices are generated, clearing concentration gradients away from the interface. While this section focussed on the data for the 0.5 A/cm$^{2}$ case, the oscillatory behaviour in the flow and concentration profiles can be observed at all current densities (see Fig. 5). At 0.3 A/cm$^{2}$ the vortices were smaller and confined to a smaller central region in the cell. At 1 A/cm$^{2}$, and in the large capacity cell, the vortices were generated at a higher frequency owing to the increased turbulence of the system.
\subsection{Prototyping and Testing}
To test the feasibility of the design, a lab-scale prototype was constructed. This allowed several aspects of the design to be evaluated, such as whether the coil can be easily electrically insulated from the liquid metal and whether the magnitudes of the flow found in the numerical model are reproduced in an experiment. Using the method detailed in section 3.3, PIV was carried out to quantitatively measure the flow induced by the coil.
\par
\begin{figure}[htp]
    \centering
    \includegraphics[width=16cm]{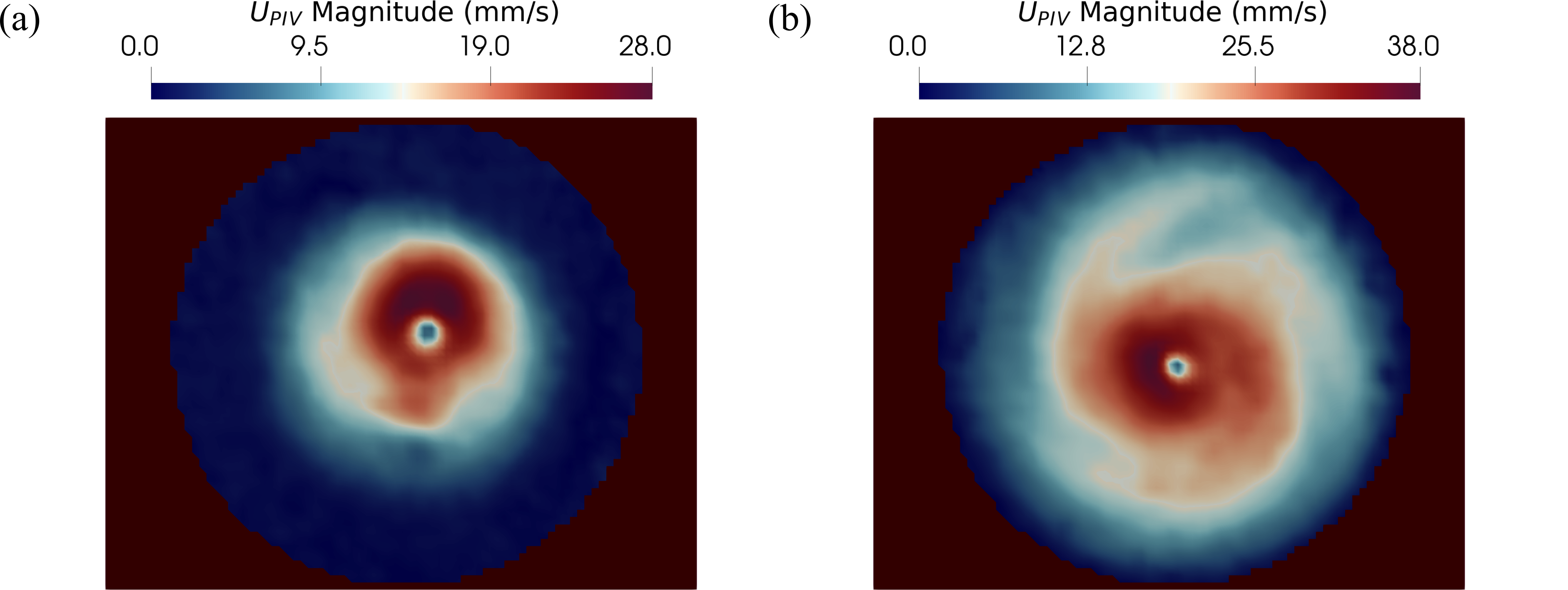}
    \caption{PIV results from the lab-scale prototype. (a) \& (b) Instantaneous snapshot of the velocity during the transient start-up and after velocity saturation}
    \label{fig:exp}
\end{figure}
For each experimental case, a 32 A current, equivalent to the discharge current in the 0.5 A/cm$^{2}$ case, was supplied to the coil, which was electrically connected in series to the current supply. The coil and current collector were implanted in a layer of Galinstan creating a flow that was initially axisymmetric as shown in Fig. 7 (a). The flow was concentrated in the central region confined by the coil and rotated in a counter-clockwise direction with a velocity of 28 mm/s. As the velocity in the layer accelerated the symmetry of the flow was broken, as shown in Fig. 7 (b). This break in the symmetry of the flow was caused by the flow achieving an azimuthal wave number of ${m=2}$ as is evident by the spiral structure in Fig. 7.
\par
Throughout each experimental run, it was observed that the vortex core was unsteady, and its position in the vessel oscillated. The vortex tracking method of Graftieaux \textit{et al.} \cite{Graftieaux2001CombiningFlows}, which uses proper orthogonal decomposition to remove noise in the PIV results and a vortex tracking algorithm to locate the vortex core, was used to track the vortex core orientation in the vessel. In Fig. 8 (a), the orientation angle with respect to the experimental vessel is displayed, and in Fig. 8 (b), the temporal evolution of the orientation angle of the vortex core is shown. During the experiment, it was found that the eye of the vortex precessed counter-clockwise as shown by the transition of the vortex core from an angle of orientation of ${\pi}$ to ${-\pi}$, which matched the direction of the flow. This precession is likely caused by the displacement of the vortex core from the centre due to turbulent fluctuations and a restoring pressure force supplied by either the coil or vessel wall. It is possible that if this motion occurred in a cathode layer it could increase the mixing in the cathode. From these results, it seems reasonable that a similar setup could be achieved in a LMB.
\par
\begin{figure}[htp]
    \centering
    \includegraphics[width=16cm]{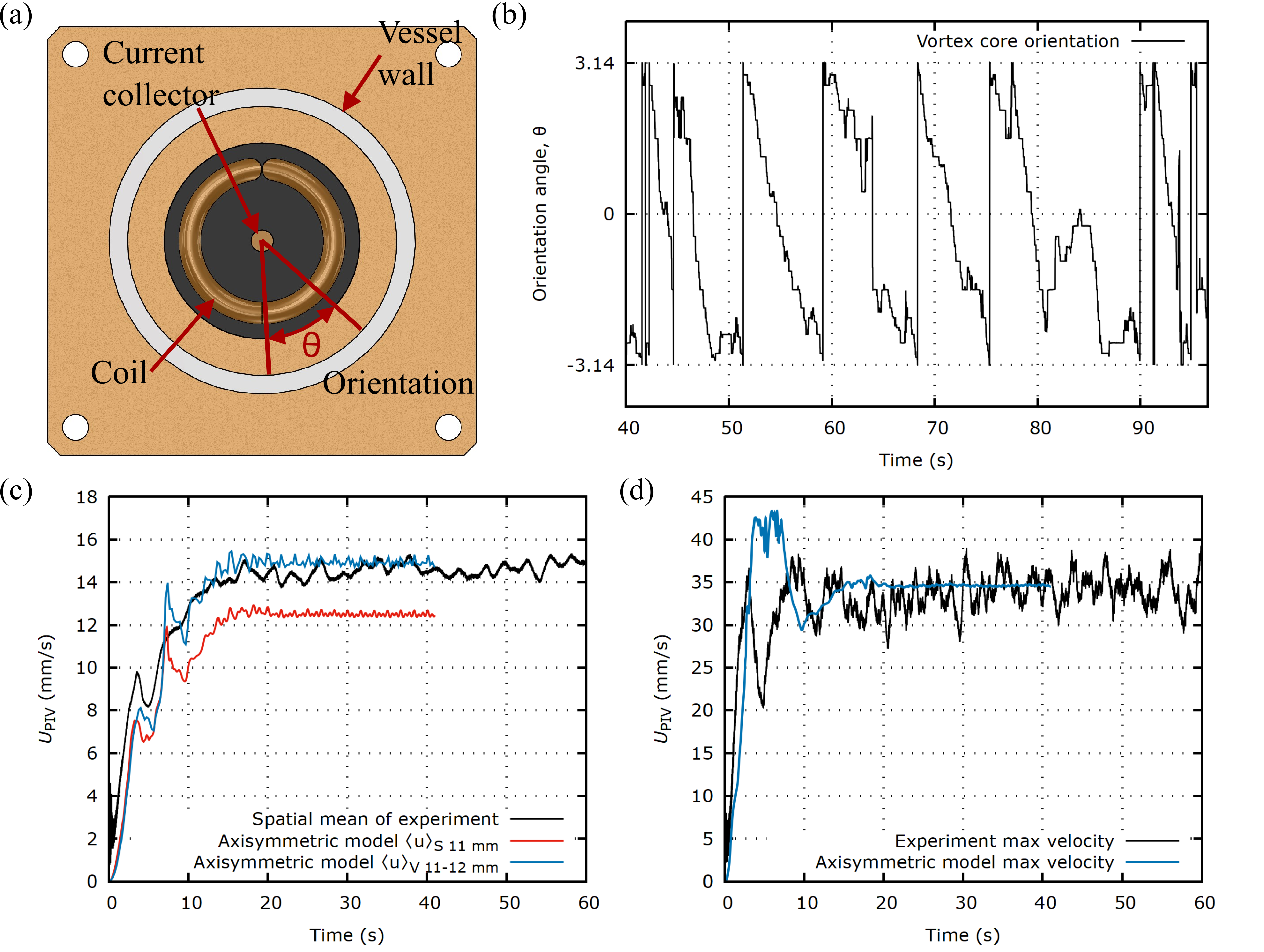}
    \caption{Time-dependent behaviour of the experimental flow. (a) Depiction of the orientation angle of the vortex core with respect to the vessel. (b) Temporal evolution of the orientation angle of the vortex core. (c) Experimental averaged velocity and numerical data. (d) Experimental maximum velocity and numerical maximum velocity}
    \label{fig:tempexp}
\end{figure}
An axisymmetric model of the experiment was computed using the geometry presented in Fig. 3. In Fig. 8 (c) \& (d), the surface averaged velocity and maximum velocity from the experiment and simulation are presented. In the figures, the velocity in the experimental case fluctuates significantly more than the data from the axisymmetric model showing that the numerical data is much steadier than the PIV data. 2D axisymmetry forces the flow to be symmetric when it may not be in a 3D model. This can cause the flow in the 2D model to have laminar boundary layers at a higher Reynolds number than when they transition to turbulence in 3D models \cite{Benard2022NumericalCylinders}. There are also additional non-axisymmetric components to the model in the lab, including the way the coil is mounted and the Earth’s background magnetic field, which can have a large effect on the magnetically driven flow \cite{Ashour2018CompetingBatteries, Keogh2021ModellingBatteries, Weber2018ElectromagneticallyBatteries}.
\par
However, axisymmetric models can still accurately capture the magnitude of the flow induced by a swirling flow, even at high current densities and magnetic field strengths \cite{Herreman2021EfficientBatteries}. To generate data from the numerical model that matched the experiment, a slice was taken at ${h=11}$ mm and the average was taken using a surface integral. It was found that at this height in the electrolyte layer, the average velocity was 12.5 mm/s which was lower than that found in the experiment, which oscillated between 14 mm/s and 16 mm/s.  Despite the low aperture used, it seems likely that a perfect slice in the electrolyte was not captured but rather a volume. A volume average of the flow from \textit{h} = 11 - 12 mm in the electrolyte layer yielded a much closer match to the experimental data, with the average from the experiment oscillating around 15 mm/s. Comparing the temporally averaged values between the numerical model and the experiment, the average values were 14.92 and 14.77 mm/s, respectively, a difference of only 1${\%}$.
\par
The maximum velocity at ${h=11}$\ mm compared much more favourably to the experiment. A weighted average filter was used with a window size of 25 values. This corresponded to the size of the interrogation windows used to analyse the flow. It was found that the maximum velocity in the simulations remained steady at 35 mm/s. The maximum velocity in the experiment oscillated over a wide range of values with velocities from 30 to 40 mm/s. The oscillations in the experiment were caused by the unsteady three-dimensional nature of the flow. The temporal mean of the max velocity from the numerical model was 34.6 mm/s, while for the experiment it was 34.3 mm/s, a difference of 0.86 ${\%}$. These results show that the numerical model used in this work captures the magnitudes of the flow well.
\subsection{Battery module concept}
\begin{figure}[htp]
    \centering
    \includegraphics[width=14.5cm]{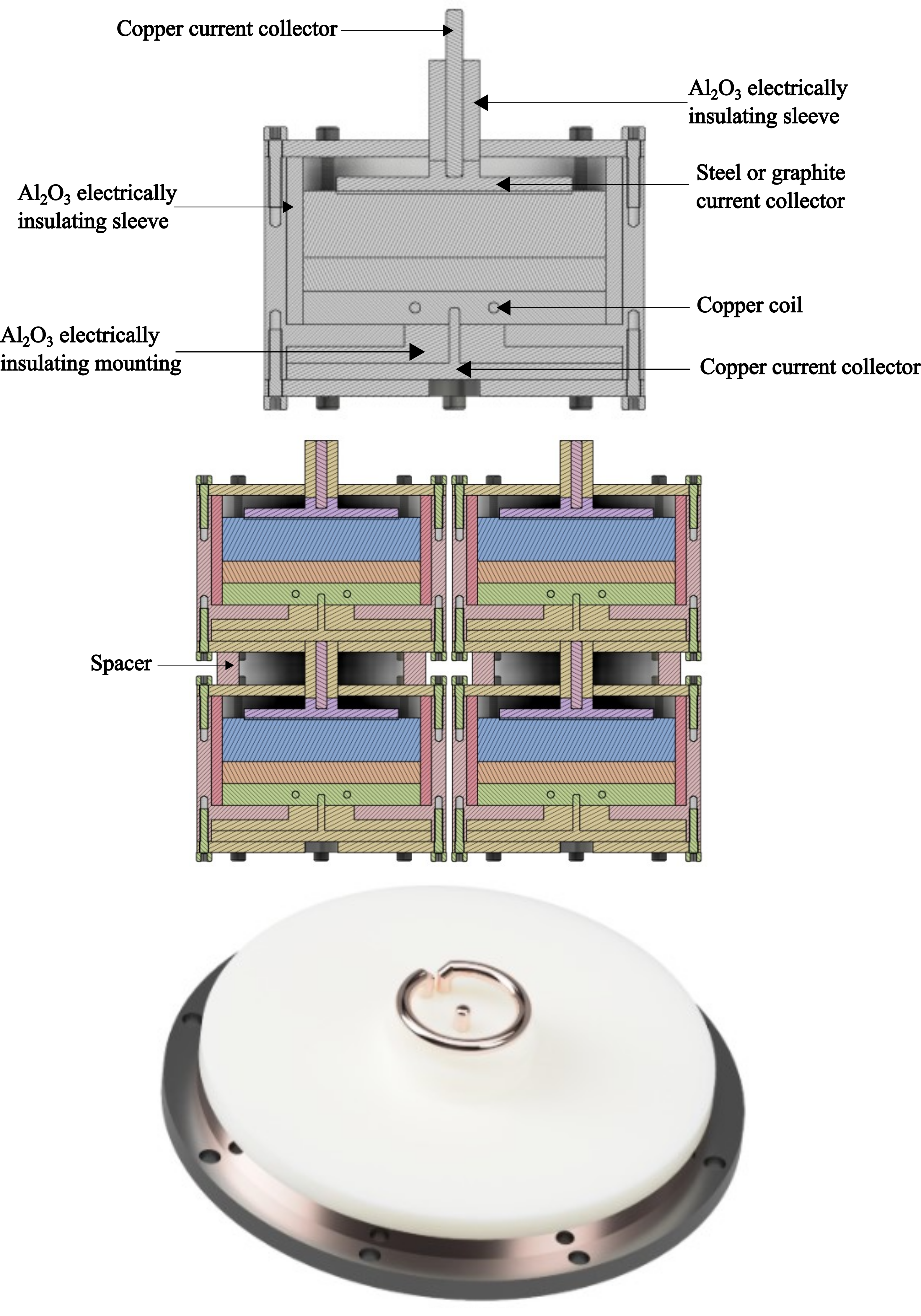}
    \caption{Rendering of a cell module concept with the coil implanted in the cathode and the modified design incorporated into a cell stack. Colours do not indicate materials.}
    \label{fig:design}
\end{figure}
\par
Possible concerns for the application of this device in LMBs include the cost, ease of manufacture, and longevity of the insulating coating that could lead to a short circuit. In Fig. 9, we seek to address these concerns by presenting a battery module and cell stack concept. In the concept, several components remain similar to current designs. The layers are still electrically isolated from the sidewalls through the use of an Al$_\mathrm{2}$O$_\mathrm{3}$ hollow cylindrical sleeve, the battery vessel is made from grade 304 stainless steel, the current collectors between vertically adjacent cells are coupled, allowing the modules to be vertically stacked and the vessels can still make electrical contact through the sidewalls allowing them to operate in parallel. By maintaining these components of current designs, existing production lines and cell construction methods can be leveraged.
\par
The coil and copper current supply are connected through a plate included in the base of the LMB. The plate connects the current supply to the coil through a circuit that is physically cast into the plate, and the plate is isolated from the cathode by a ceramic layer. To minimise additional manufacturing costs, the current supply and coil could be manufactured at scale using sand or die casting. Both methods offer efficiency at scale and are particularly suited to mass production for parts made using copper. This would allow for the low cell costs to be maintained. With the gains to the power density of LMBs utilising this coil, there should be no increase to the costs per kilowatt hour. 
\par
The coil and current supply could foreseeably become a wear item in the cell as other works in DC electric arc furnaces have shown that electrodes exposed to liquid metal flows can be worn away over time \cite{Kazak2013ModelingPositions}. The ceramic coating, however, is identical to the insulation currently used in LMBs. It should be reasonable to expect that the batteries will maintain the same lifespan regardless of this modification. 
\par
If a coil fails or begins to short circuit by bolting on the top and bottom lids, the internal components of the cell remain accessible. A particularly advantageous aspect of current LMB cell-module designs is that they are modular. If a single cell fails, it can be removed and replaced with relative ease.
\par
The analysis in this work shows that the inclusion of this design in future LMBs could lead to considerable performance and capacity gains. The next step in developing this concept should be the construction of a demonstration cell so that the increases in the electrochemical performance can be validated. 
\section{Conclusion}
In this work, a numerical model was developed of a three-layer LMB using a multi-region approach. An alternative design consisting of a coil implanted in the cathode layer to induce a mixing flow was tested to see if the electrochemical performance of the battery could be improved. Comparison against a baseline diffusion case after a ${\Delta x_\mathrm{Li}}$ of 0.1 showed that the alternative design substantially improved concentration polarisation in the cathode layer with an 81${\%}$ and 88${\%}$ reduction in the divergence of the interfacial concentration from the volumetric average concentration at 0.3 and 0.5 A/cm$^{2}$, respectively. The improved interfacial concentrations increased the discharge voltage of the battery with the voltage increasing by 8${\%}$ and 13${\%}$ after a ${\Delta x_\mathrm{Li}}$ of 0.1 for the 0.3 and 0.5 A/cm$^{2}$ cases. Further, for the length of the computations in all cases, the interfacial concentration increased linearly as the volumetric average concentration increased, which suggests that a LMB utilising this design could achieve complete anode utilisation even at large discharge capacities of 481 Ah, which was modelled in this work. The results were analysed to see how the concentration profiles were affected by different discharge current densities, and it was found that the rate of induced mixing in the cathode increased faster than the rate of mass flux. This means that no mass-transport penalty was experienced by the modified LMB at higher discharge current densities.
\par
The improved performance of the battery was attributed to the flow in the cathode. The coil induced a vigorous swirl flow in the cathode, but the geometrical configuration of the cathode layer substantially changed the dynamics of flow as they have been described in past works. Rather than the dynamics being dominated by the toroidal velocity, it was found that the coil substantially increased the poloidal velocity in the layer such that it dominated the dynamics of the flow. The reduced interfacial concentrations were caused by the periodic formation of vortices that washed away lithium atoms from the cathode-electrolyte interface. These vortices displayed temporal coherence with peak frequencies of 0.0637 and 0.0771 Hz.
\par
The feasibility and viability of the design was tested by construction of a lab scale prototype. The prototype was tested in a layer of Galinstan using PIV. A thin, seeded layer of HCl was used to dissolve the oxide layer on the Galinstan, allowing velocity to be transferred to the HCl via viscous coupling.  The results from the experiment closely matched the numerical results validating the modelling methodology used. This design concept could lead to the realisation of LMBs with both extended discharge capacities and high discharge voltages. 
\section{Declaration of interests}
Declan Finn Keogh has patent Novel electrode concept for performance improvement in liquid metal batteries issued to UNSW. Chris Menictas has patent Novel electrode concept for performance improvement in liquid metal batteries pending to UNSW. Victoria Timchenko has patent Novel electrode concept for performance improvement in liquid metal batteries pending to UNSW. Mark Baldry has patent Novel electrode concept for performance improvement in liquid metal batteries pending to UNSW. If there are other authors, they declare that they have no known competing financial interests or personal relationships that could have appeared to influence the work reported in this paper.
\section{Acknowledgements}
This work was carried out with the support of an Australian Government Research Training Program (RTP) Scholarship. Computations were performed on the super-computer Gadi using a grant awarded under NCMAS-2022-69 with the assistance of resources and services from the National Computational Infrastructure (NCI), which is supported by the Australian Government. The authors would like to thank Dr Charitha de Silva for assisting with the PIV method.
\




 
\bibliography{references}


\end{document}